\documentclass[12pt]{article}
\usepackage{latexsym}

\def\bequ{\begin{equation}}
\def\eequ{\end{equation}}
\def\bea{\begin{eqnarray}}
\def\eea{\end{eqnarray}}
\def\barr{\begin{array}}
\def\earr{\end{array}}
\def\bfone{\relax{\rm 1\kern-.35em 1}}

\def\IE{\relax{{\rm I\kern-.18em E}}}
\def\IGam{\relax{{\rm I}\kern-.18em \Gamma}}

\def\IA{\relax{\hbox{{\rm A}\kern-.82em {\rm A}}}}

\def\o#1#2{{{#1}\over{#2}}}

\def\IP{\relax{\rm I\kern-.18em P}}

\font\cmss=cmss10 \font\cmsss=cmss10 at 7pt

\def\inbar{\vrule height1.5ex width.4pt depth0pt}
\def\IC{\relax\,\hbox{$\inbar\kern-.3em{\rm C}$}}
\def\IG{\relax\,\hbox{$\inbar\kern-.3em{\rm G}$}}
\def\IB{\relax{\rm I\kern-.18em B}}
\def\ID{\relax{\rm I\kern-.18em D}}
\def\IL{\relax{\rm I\kern-.18em L}}
\def\IE{\relax{\rm I\kern-.18em E}}
\def\IF{\relax{\rm I\kern-.18em F}}
\def\IH{\relax{\rm I\kern-.18em H}}
\def\II{\relax{\rm I\kern-.17em I}}
\def\IN{\relax{\rm I\kern-.18em N}}
\def\IP{\relax{\rm I\kern-.18em P}}
\def\IQ{\relax\,\hbox{$\inbar\kern-.3em{\rm Q}$}}
\def\bfzero{\relax\,\hbox{$\inbar\kern-.3em{\rm 0}$}}
\def\IR{\relax{\rm I\kern-.18em R}}
\def\ZZ{\relax\ifmmode\mathchoice
{\hbox{\cmss Z\kern-.4em Z}}{\hbox{\cmss Z\kern-.4em Z}}
{\lower.9pt\hbox{\cmsss Z\kern-.4em Z}}
{\lower1.2pt\hbox{\cmsss Z\kern-.4em Z}}\else{\cmss Z\kern-.4em
Z}\fi}
\def\IU{\relax\,\hbox{$\inbar\kern-.3em{\rm U}$}}
\def\bfone{\relax{\rm 1\kern-.35em 1}}

\def\trace{{\rm Tr}\hskip 1pt}
\def\o#1#2{{{#1}\over{#2}}}
\def\eq#1{(\ref{#1})}
\def\ii{{\rm i}}

\def\bar{\overline}

\def\Coe#1.#2.{{#1\over #2}}

\def\coe#1.#2.{\relax{\textstyle {#1 \over #2}}\displaystyle}

\def\to{\rightarrow}
\def\notin{\hbox{{$\in$}\kern-.51em\hbox{/}}}
\def\notdel{\hbox{{$\partial$}\kern-.51em\hbox{$\backslash$ }}}

\def\ie{{\it i.e.}}
\newcommand{\be}{\begin{equation}}
\newcommand{\ee}{\end{equation}}

\csname @addtoreset\endcsname{equation}{section}
\newcommand{\nn}{\nonumber}\newcommand{\p}[1]{(\ref{#1})}
 \newcommand{\lb}[1]{\label{#1}}
\newcommand\s{\scriptscriptstyle}

\newcommand\q{\quad}
\newcommand\qq{\qquad}
\newcommand{\bbox}{\lower.2ex\hbox{$\Box$}}

\newcommand\tia{\theta^\alpha_i}

\newcommand\btia{\bar{\theta}^{\da i}}

\newcommand{\da}{{\dot{\alpha}}}
\newcommand{\db}{{\dot{\beta}}}

\newcommand{\toa}{\theta_1^\alpha}

\newcommand{\tta}{\theta_2^\alpha}

\newcommand{\tha}{\theta_3^\alpha}

\newcommand{\tfa}{\theta_4^\alpha}

\newcommand{\bto}{\bar{\theta}^1}
\newcommand{\btoa}{\bar{\theta}^{1\da}}
\newcommand{\btob}{\bar{\theta}^{1\db}}
\newcommand{\btt}{\bar{\theta}^2}
\newcommand{\btta}{\bar{\theta}^{2\da}}
\newcommand{\bttb}{\bar{\theta}^{2\db}}
\newcommand{\bth}{\bar{\theta}^3}
\newcommand{\btha}{\bar{\theta}^{3\da}}
\newcommand{\bthb}{\bar{\theta}^{3\db}}

\newcommand{\btfa}{\bar{\theta}^{4\da}}

\newcommand\ada{{\alpha\da}}
\newcommand\adb{{\alpha\db}}

\newcommand\padb{\partial_\adb}
\newcommand\pada{\partial_\ada}

\newcommand\A{{\s A}}

\newcommand{\Dot}{D^1_2}
\newcommand{\Dto}{D^2_1}
\newcommand{\Doh}{D^1_3}

\newcommand{\Dth}{D^2_3}

\newcommand{\poh}{\partial^1_3}

\newcommand{\poo}{\partial^1_1}
\newcommand{\ptt}{\partial^2_2}
\newcommand{\phh}{\partial^3_3}

\newcommand{\pff}{\partial^4_4}

\newcommand{\Doa}{D^1_\alpha}

\newcommand{\Dta}{D^2_\alpha}

\newcommand{\Dha}{D^3_\alpha}

\newcommand{\poa}{\partial^1_\alpha}

\newcommand{\bDoa}{\bar{D}_{1\da}}

\newcommand{\bDta}{\bar{D}_{2\da}}
\newcommand{\bDtb}{\bar{D}_{2\db}}
\newcommand{\bDha}{\bar{D}_{3\da}}

\newcommand{\bDfa}{\bar{D}_{4\da}}

\newcommand{\bpha}{\bar{\partial}_{3\da}}

\begin{document}

\begin{titlepage}
\begin{flushright}
CERN-TH/99-349\\ KUL-TF-99/39\\ LAPTH-766/99\\JINR-E2-99-309 
\end{flushright}
\vspace{.5cm}
\begin{center}
{\Large\bf Shortening of primary operators in $N$-extended
SCFT$_4$ and harmonic-superspace analyticity}\\
\vfill {\large L. Andrianopoli$^1$, S. Ferrara$^2$,
 E. Sokatchev$^3$, B. Zupnik$^4$  }\\
\vfill

{\small
$^1$ Instituut voor Theoretische Fysica, Katholieke
 Universiteit Leuven,\\
Celestijnenlaan 200D B-3001 Leuven, Belgium
\\ \vspace{6pt}
$^2$ CERN Theoretical Division, CH 1211 Geneva 23, Switzerland
\\ \vspace{6pt}
$^3$ Laboratoire d'Annecy-le-Vieux de Physique
Th\'{e}orique\footnote[5]{UMR 5108 associ{\'e}e {\`a}
 l'Universit{\'e} de Savoie} LAPTH, Chemin
de Bellevue - BP 110 - F-74941 Annecy-le-Vieux Cedex, France
\\ \vspace{6pt}
$^4$ Bogoliubov Laboratory of Theoretical Physics, Joint Institute for
Nuclear Research, Dubna, Moscow Region, 141980, Russia
}

\end{center}
\vfill

\begin{center}
{\bf Abstract}
\end{center}
{\small We present the analysis of all possible shortenings which
occur for composite gauge invariant conformal primary superfields
in $SU(2,2/N)$ invariant gauge theories.\\ These primaries have
top-spin range $\o N2 \leq J_{max} < N$ with $J_{max} =J_1+J_2$,
($J_1,J_2$) being the $SL(2,\IC)$ quantum numbers of the highest
spin component of the superfield.\\ In Harmonic superspace,
analytic and chiral superfields give $J_{max}=\o N2$ series while
intermediate shortenings correspond to fusion of chiral with
analytic in $N=2$, or analytic with different analytic structures
in $N=3,4$.\\In the AdS/CFT language shortenings of UIR's
correspond to all possible BPS conditions on bulk states.
\\ An application of this analysis to multitrace
operators, corresponding to multiparticle supergravity states, is
spelled out. }
\end{titlepage}
\section{Introduction}
 The recent
interplay between supergravity in $AdS_5$ and superconformal
 $SU({\mathcal N})$
Yang--Mills theories in the large ${\mathcal N}$ limit
\cite{mal,gkp,wit}
 has lead to a deeper
investigation of $SU(2,2/N)$ superconformal algebras and their UIR's
both on bulk states and on superfield boundary operators.

A complete identification of highest weight UIR's was given in ref.
\cite{ff} for $N=1$ and further extended to any $N$ by Dobrev and
Petkova \cite{dp} and also in \cite{bin,mss}.\\
Since Yang--Mills theories are built only with a finite number
of supersingleton fields, having $J_{max} \leq 1$ (these are the basic
multiplets of the 4 dimensional superconformal theory) only a subclass
of all possible UIR's are realized in QFT, nevertheless the variety of such
representations is still rather rich and many different shortenings
may occur.

Short multiplets have an important aspect in the $AdS/CFT$
correspondence because they have ``protected'' conformal
dimensions and therefore allow a reliable comparison between
quantities computed in the bulk versus quantities derived in the
$CFT_4$ \cite{So,fpz}.

A particular example of such a phenomenon are the K--K masses of
bulk states which belong to short $SU(2,2/N)$ UIR's \cite{gm}. For
such states in the $N=4$ case, corresponding to IIB supergravity
on $AdS_5 \times S_5$ \cite{gurw},
 it is possible to give, at least for large ${\mathcal N}$, an ``exact''
operator realization in terms of 4d shortened superconformal fields
\cite{fefrza,AF}.

Another example of such a correspondence has been worked out in
the literature \cite{gu,cddf} by comparing IIB supergravity on
$AdS_5 \times T_{11}$ \cite{ro} and a specific $SU(2,2/1)$
invariant $SU({\mathcal N})\times SU({\mathcal N})$
Yang--Mills theory constructed by Klebanov and Witten
\cite{kw}.

Already in this simple $N=1$ example it was realized that $N=1$ 
chiral superfields \cite{fwz} are only a particular case of short 
representations. Indeed it was shown \cite{cddf} that semishort 
multiplets occur in the K--K spectrum of IIB supergravity on 
$AdS_5 \times T_{11}$, with the very subtle implication that some 
square root formulae for the conformal dimensions, giving in 
general irrational numbers, become perfect squares for particular 
relations of the quantum numbers of the bulk states, precisely 
corresponding to semi-shortening conditions, which imply rational 
conformal dimensions. 

For $N>1$ $SCFT_4$'s the shortening and semi-shortening is even
richer because maximal shortening (which means half of the total
number of $\theta$'s) can occur either with chiral superfields or
with ``Grassmann (G--)analytic'' superfields 
\cite{GIK01}-\cite{hh1}. 

This may happen because a new class of UIR's occur for $N\geq 2$
which have no $N=1$ analogue (class C) in the classification given
below \cite{dp}).

The $N=2$ hypermultiplets and the $N=3,4$ Yang--Mills field 
strength multiplets (supersingletons) belong precisely to this 
class of UIR's together with an infinite tower of recurrences. At 
the same, in superspace they are described by the new type of 
short (G--analytic) superfields (see \cite{GIK01}-\cite{GIK1} for 
$N=2$ and \cite{hh,hh1} for $N=3,4$).  

A crucial ingredient to understand the occurrence of different 
shortenings for composite superconformal primaries in $N= 2,3,4$ 
theories is the use of harmonic superspaces 
\cite{GIK1,GIO,IKNO,hh} with harmonic variables on 
$\o{SU(N)}{U(1)^{N-1}}\;$, {\ie} the coset given by non-Abelian 
R--symmetry modded by its Cartan subalgebra (the maximal torus). 

Analytic superfields (in harmonic superspace) correspond to a new
class of $N\geq 2$ UIR's which have no $N=1$ analogue. Moreover
for $N>2$, since the above coset has many complex variables,
different types of analyticity may occur and this allows for an
even richer structure of shortenings when composites of
superfields with different G--analytic structure are considered.

\vskip 5mm
The present paper is organized as follows:

In section 2 we review the unitarity bounds of highest weight
UIR's of $SU(2,2/N)$ ($N\geq 2$) and of $PSU(2,2/4)$ for $N=4$.

In section 3 we consider extended superspaces with harmonic
variables in $\o{SU(N)}{U(1)^{N-1}}$ for $N=2,3,4$ in subsections 1, 2, 3
respectively. G--analytic
properties of ``supersingleton'' representations (the would-be
massless fields on the boundary) are explained together with
different analytic structures occurring for $N=2,3,4$.
\\
Note that although $N=3$ Yang--Mills is believed to be the same of $N=4$,
certainly this is not the case for the bulk theory since $N=6$ supergravity on
$AdS_5$ is not the same as $N=8$ \cite{grw}.
\\
This can be understood from the fact that there is a ring of operators which
reproduces the $N=6$ supergravity states \cite{fpz,is}.
\\
Another fact is that $N=3$ harmonic superspace provides an
off-shell formulation of maximally supersymmetric Yang--Mills
theory which is not available in its own $N=4$ superspace
\cite{GIO,kall}.
\\
The three subsections are written in an independent way, so that the reader
not interested can skip any of them.

In section 4 we consider an application to multitrace operators in
$N=4$ Yang--Mills theory by showing that all such operators have some
superconformal irreducible components which are short with
different types of shortenings.
\\
In  the $AdS_5/CFT_4$ correspondence they should correspond to
multiparticle BPS supergravity states preserving respectively $\o
12$, $\o 14$ and $\o 18$ supersymmetries. The first two types of
states occur in the double-trace operators while the third type
starts to occur for triple-trace operators.

In section 5 a summary of our results is presented.
Some technical material related to properties of harmonic superspace
is collected in an appendix.

\setcounter{equation}0

\section{Highest weight UIR's of $SU(2,2/N)$: a review }
The $SU(2,2/N)$ superalgebra has a 5-grading decomposition
\cite{bgg,gmz1,gmz2}:
\be
{\mathcal L}_N= {\mathcal L}^1 + {\mathcal L}^{\o 12} + {\mathcal 
L}^0 + {\mathcal L}^{-\o 12} + {\mathcal L}^{-1} \ee with respect 
to its maximal compact subalgebra ${\mathcal L}^0=SU(2) \times 
SU(2) \times U(1) \times U(N)$. 

A highest weight state is defined as \be {\mathcal L}^{-\o
12}|\Omega > = {\mathcal L}^{-1}|\Omega >=0\;. \ee A highest
weight UIR representation is specified by a UIR rep. of $\Omega$
with respect to ${\mathcal L}^0$. The eigenvalues of the two
$U(1)$ respectively denote $E_0$ (the AdS energy) and the $U(1)$
R-charge.

The ``compact basis'' is suitable to discuss bulk states, {\ie}
UIR's  on $AdS_5$. In the AdS/CFT correspondence the CFT operators
are naturally described in the ``non-compact basis''\cite{gmz2},
in which the highest weight state is mapped into a space--time
superfield whose lowest component $\phi (x,\theta )|_{x=\theta
=0}$ is a irreducible representation of $SL(2,\IC )\times
O(1,1)\times U(N)$.

In this correspondence the ($J_1 , J_2 $) quantum numbers denote a
$SL(2,\IC )$ irreducible representation and $E_0 \to \ell$
 denotes the dilatational
 weight (conformal dimension).

A space--time superfield, whose lowest component corresponds to a highest
weight state, is called a ``primary'' (or quasi-primary) conformal superfield.
Needless to say that $SU(2,2/N)$ will have ``supercasimir'' operators
which are at least $3+N= $ rank$(SU(2,2)\times U(N))$ ($3+N-1$ for $N=4$ if we
consider the superalgebra 
$PSU(2,2/4)$ \cite{bin} as we will do in the present paper).

In what follows it will be convenient to define ``Poincar\'e''
supercasimirs for the $SU(N)$ part of the superconformal algebra
in order to study the irreducible content of the superfields in
Harmonic superspace (HSS). 
The other quantum number properties will be straightforward.

Since a UIR of $SU(2,2/N)$ is denoted by its highest weight state,
we will mainly denote such rep. \cite{ff} by ${\mathcal D}(\ell ,
J_1,J_2;r;a_1,\ldots ,a_{N-1})$, where $a_1,\ldots ,a_{N-1}$ are
the Dynkin labels of $SU(N)$. For a given $N$ we call $a_1 =a$,
$a_2 =b$, $a_3 =c, \ldots$. Here $\ell$ is the conformal dimension
of $\phi(x=0, \theta =0)$ and $(J_1,J_2,r)$ its $SL(2,\IC)$ and
$U(1)$ R-symmetry quantum numbers.

In the next sections we will only consider the cases with $J_1=J_2=0$
and supersingletons with top spin $J\leq 1$.

The quantum numbers of the highest weight state are subjected to
some unitarity bounds \cite{dp}, whose thresholds correspond to the several
possible shortenings of UIR's of $SU(2,2/N)$.

When the maximal  possible number of bounds are fulfilled then the UIR becomes
extrashort, in the sense that it gives the least possible number of
states for a given set of unitarity bounds.

Examples of such extrashort UIR's are the ``supersingletons'' and
the bulk ``massless'' reps., which in the CFT language correspond
to boundary massless fields and to conserved current operators respectively
\cite{fefrza,FZ}.

Supersingletons \cite{ff2} are called ``ultrashort'' because their 
degrees of freedom are not enough to correspond to particle states on AdS
bulk.

On these UIR's space-time derivative constraints are imposed on
the conformal primary operators. For all other shortenings no
space-time derivative constraints are imposed but rather a
relation between different $\theta$ components of the conformal
superfield at hand.

\vskip 5mm

The highest weight UIR's of the superconformal algebras
$SU(2,2/N)$ fall in three categories, depending on the quantum
numbers of the highest weight state
\be
{\mathcal D}(\ell ,J_1,J_2;r,m_1,\ldots , m_{N-1}) \ee where
$\ell$ and $(J_1,J_2)$ label the dimension and the $SL(2,\IC)$
spin respectively, $r$ is the $U(1)$ R--symmetry and $(m_1,\ldots
, m_{N-1})$ the Young tableaux (YT) labels of $SU(N)$ (
$m=\sum_{k=1}^{N-1} m_k$).

Let us define the quantities:
\begin{eqnarray}
X(J,r,2\frac{m}{N}) &=&2+2J-r+ \frac{2m}{N}
\nonumber\\
Y(r, 2\frac{m}{N}) &=&-r+ \frac{2m}{N}
\end{eqnarray}
Then we have \cite{dp} ($J_1=J_L$, $J_2=J_R$):
\begin{itemize}
\item{
A) $\ell\geq X(J_2,r,2\frac{m}{N})
 \geq X(J_1,-r, 2m_1 -2\frac{m}{N})$,\\
(or $J_1 \to J_2$, $r\to -r$, $2\frac{m}{N}\to  2m_1 - 2\frac{m}{N}$),
 for $J_1J_2 \geq 0$.
}
\item{
B) $\ell =Y(r, 2\frac{m}{N}) \geq X(J_1,-r,2m_1- 2\frac{m}{N})$,\\
(or $J_1 \to J_2$, $r\to -r$, $ 2\frac{m}{N} \to 2m_1 - 2\frac{m}{N}$),
 for $J_1J_2= 0$, $J_1=J$.
}
\item{
C) $\ell =m_1$, $r=2\frac{m}{N}-m_1$, $J_1=J_2=0$
}
\end{itemize}

``Massless'' representations in the $AdS_5$ bulk correspond to the 
threshold in A) when $\ell = 2+J_1+J_2$, $r=-J_1+ J_2$, in B) when 
$\ell =-r=2+J$ and in C) when $\ell =m_1=2$. 

In these cases the CFT superfield is such that ``current'' components
of the form
$J_{\alpha_1\cdots\alpha_{2s_1},\dot\alpha_1\cdots\dot\alpha_{2s_2}} $,
with $\ell_J=2+s_1+s_2$, are conserved:
\be
\partial^{\alpha_1\dot\alpha_1}
J_{\alpha_1\cdots\alpha_{2s_1},\dot\alpha_1\cdots\dot\alpha_{2s_2}}=0
\ee
We will call the superfield in question ``current superfield''.
The ``supersingletons'' (massless conformal fields) occur in B) for $\ell
=-r=1+J$ and in C) for $\ell =m_1 =1$.
These representations are ``ultrashort'' and the field components
$O_{\alpha_1\cdots\alpha_{2s}}$ obey the equatons of motion:
\be
\partial^{\alpha_1\dot\alpha_1}
O_{\alpha_1\cdots\alpha_{2s}}=0 \qquad ( \bbox ~O=0 \mbox{ for } s=0).
\ee

It is a general fact of AdS reps. that ``massless bulk'' UIR's are
contained in the product of two singletons \cite{ff2}.

In the SCFT language this means that a ``current superfield'' is bilinear
in the supersingleton superfields as one can easily check.

Other shortenings, which do not involve space-time constraints on component
fields, will be called ``short multiplets'' or ``short superfields'' where
the shortening has a ring structure, namely it is preserved by multiplication
of superfields.

As we will see in the next section, this corresponds to the concept
of ``chirality'' and Grassmann analyticity in Harmonic superspace.
\\
Shortenings of this type occur in B) and C).

Finally, there are other types of shortenings which are not of this type;
we will call them ``semishort''.
They can appear in A) B) C) and typically correspond to superfields
which satisfy second or higher order constraints in covariant derivatives
(second order constraints define the so-called linear superfields).
In all the above cases no space-time constraints on component fields
are implied.

Here we will consider the cases $N=2,3,4$; the $N=1$ case has been
treated elsewhere \cite{ff,fgp,osb}.

We shall use Dynkin labels (DL) $[a_1,\ldots , a_{n-1}]$ for
$SU(n)$, which are related to the YT labels $(m_1,\ldots ,
m_{n-1})$ as follows:
\be
m_i=\sum_{k=1}^{n-i} a_k\; , \quad i=1,\ldots , n-1\; . \ee

A crucial ingredient in our analysis will be the use of harmonic
superspace $(x,\theta , u)$, with ``harmonic'' variables $u$
parametrizing the coset ${SU(N)}/{U(1)^{N-1}}$.

We will separately consider the cases $N=2,3,4$ and for $N=4$ we will restrict
the analysis to the $PSU(2,2/4)$ algebra ($r=0$) \cite{dp,bin},
 since it is the latter which
is appropriate to $N=4$ super Yang--Mills theory.

\setcounter{equation}0


\section{Extended harmonic superspaces and short superfields}

We are interested in realizing the highest weight UIR's of
$SU(2,2/N)$ on superfields and harmonic superfields.

We shall consider an $N$--extended $D=4$ superspace (without
central charges) with Grassmann coordinates manifestly covariant
with respect to the $SU(N)$ group
\begin{equation}
z=(x^\adb ,~\tia,~\btia) \lb{B1} \ee using spinor indices of
$SL(2,C)$ and indices in the  fundamental representations of
$SU(N)~ i, k, \ldots=1,\ldots, N$. (Note that the alternative
convention $\theta^{\alpha i},~\bar{\theta}^\da_i$ is sometimes
used in the literature.)

The covariant opeartors in superspace are the spinor derivatives
(see (\ref{a16})). Using them one can define constrained
superfields describing various irreducible representations of
extended supersymmetries. The construction of supercurrents and
superactions from the constrained superfields have been discussed
in Refs. \cite{So,HST}.

For $N>1$ the standard superspace (\ref{B1}) can be enlarged to a 
``harmonic superspace" (HSS) \cite{GIK1} by considering an extra 
set of ``harmonic coordinates'' $u$  which provide an 
$SU(N)$-covariant parametrization of the coset space 
\be
{\mathcal M} = \frac{SU(N)}{S\left(U(n_1)\times\ldots\times
U(n_P)\right)}\;, \qq \left(\sum_{k=1}^P n_k =N\right). \lb{flag} 
\ee Note that such spaces are known in the mathematical literature 
under the name of ``flag manifolds'' \cite{Ro,hh1,knapp}. An 
exhaustive list of these space and the corresponding HSS's and of 
their properties for $N=2,3,4$ is given in Ref. \cite{IKNO}. The 
choice of the subgroup depends on the practical use made of the 
harmonic variables. In our context it is crucial to use the 
highest-dimensional manifold of type \eq{flag} which occurs by 
dividing $SU(N)$ by its Maximal Torus: 
\be
{\mathcal M_C} = \frac{SU(N)}{\left(U(1)\right)^{N-1}}\;.
\lb{flagcartan} \ee
 This is a
manifold of complex dimension ${N(N-1)}/{2}$. The advantage of
this choice is that the residual symmetry
$\left(U(1)\right)^{N-1}$ is the smallest one possible. This gives
us maximal flexibility in defining subspaces of the full
$N$--extended superspace in an $SU(N)$ covariant way. Such
subspaces contain only a subset of the $4N$ Grassmann variables
and are therefore called Grassmann (G--)analytic. They can be
viewed as an alternative to the familiar chiral subspaces \cite{fwz}.

The $N=2$ and $N=3$ HSS's based on the cosets $SU(2)/U(1)$ and 
$SU(3)/U(1)\times U(1)$, correspondingly, have been introduced for 
the off-shell description of all the $N=2$ supersymmetric theories 
and of $N=3$ supersymmetric Yang-Mills theory (SYM) 
\cite{GIK1}-\cite{kall}. The realization of the superconformal 
group in these HSS's has been studied in Refs. \cite{GIOS3,GIO1}. 
An $N=4$ HSS involving the manifold $SU(4)/S(U(2)\times U(2))$ has 
been used in Ref. \cite{hh} to give an interpretation of the 
on-shell constraints of $N=4$ SYM as combined G-- and harmonic 
(H--)analyticities. There it was shown that the on-shell 
superfield strength of $N=4$ SYM is an analytic (i.e., short) 
superfield (and, similarly, for $N=3$ SYM). In Ref. \cite{HW} this 
observation was generalized to composite operators made out of 
$N=4$ SYM field strenghts and it was suggested that this property 
might significantly restrict the  correlation functions of 
analytic composite operators. This idea has subsequently been 
applied to the study of $N=2$ and $N=4$ CFT's in Refs. 
\cite{EHSSW}. The notion of harmonic superspace and of 
G--analyticity was generalized to an arbitrary $N$ in 
\cite{hh,hh1}. Note that H--analyticity is also important in HSS's 
of lower space-time dimension \cite{hlee,Z4}. 

In what follows we shall consider the irreducible superfield
representation of $SU(2,2/N)$ using the G-- and H--analyticities
related to the choice (\ref{flagcartan}).

\subsection{The $N=2$ case}

$N=2$ HSS has been introduced for the off-shell description of the
hypermultiplet and gauge and supergravity multiplets \cite{GIK1}.
Here we start by describing the shortening effect of G-- and
H--analyticities on the on-shell matter and gauge $N=2$ multiplets
satisfying their free equations of motion. We review the basic
facts about the $SU(2)/U(1)$ HSS in the Appendix.

The hypermultiplet (or $N=2$ matter multiplet), which is the
supersingleton in the AdS literature, can be described by an
ordinary superfield which is an $SU(2)$ doublet $q^i(z)$ and
satisfies the on-shell constraints \cite{Sohnius}:
\be
D_\alpha^{(i} q^{j)}(z) = \bar D_{\dot\alpha}^{(i} q^{j)}(z) =0\;.
\label{2constr} \ee Now, let us project the $SU(2)$  doublets
$D_\alpha^{i}, q^i$ and $\bar D_{i\dot\alpha}$ with the harmonics
$u^1_i$ and $u^i_2$
\begin{equation}\label{prj}
  D^1_\alpha = D_\alpha^{i}u^1_i\;,\quad q^1 =q^i
u^1_i\;, \quad \bar D_{2\dot\alpha} = \bar D_{i\dot\alpha}u^i_2
\;.
\end{equation}
This allows us to equivalently rewrite the on-shell constraints
(\ref{2constr}) in the form of G--analyticity conditions in HSS:
\begin{equation}\label{C0}
  D^1_\alpha q^1 = \bar D_{2\dot\alpha}q^1=0\;.
\end{equation}

The crucial point now is that by letting the superfield $q^1$ have
a non-trivial harmonic dependence one can solve the constraints
eqs. (\ref{C0}) in terms of a G--analytic superfield
$q^1(x_\A,\theta_2,\bar{\theta}^1,u) \equiv q^+$ in the
appropriate analytic basis (\ref{a14}). This superfield describes
the hypermultiplet off shell. It is an  infinite-dimensional
representation of supersymmetry because of the infinite harmonic
expansion on the coset $SU(2)/U(1)\sim S^2$. In order to put it
back on shell we need to restrict the arbitrary harmonic
dependence down to the initial linear one (\ref{prj}). This is
achieved with the help of the harmonic derivatives defined in the
Appendix. We remark that the harmonic derivative $D^{\,1}_2$
commutes with the spinor ones from eq. (\ref{C0}),
\begin{equation}\label{C000}
  [D^{\,1}_2,D^1_\alpha]=[D^{\,1}_2,\bar D_{2\dot\alpha}]=0\;,
\end{equation}
 {\ie} preserves G--analyticity.  Thus, the
free equation of motion of the hypermultiplet takes the form of a
harmonic (H-)analyticity condition:
\be
\Dot q^1=0~,\lb{C1} \ee where one uses the harmonic derivative in
the analytic basis (see \p{a18}).

This harmonic equation implies a
number of constraints on the components of the superfield $q^1$.
Take, for instance, the leading component $\phi^1(x,u) =
q^1\vert_{\theta=0}$. Off shell it has an infinite harmonic
expansion going over the irreducible products of the harmonics
$u^{1,2}_i$:
\begin{equation}\label{infexp}
  \phi^1(x,u) = \sum_{n=0}^\infty \phi^{(i_1\ldots i_{n+1}j_1\ldots
j_n)}(x) u^1_{i_1}\ldots u^1_{i_{n+1}} u^2_{j_1}\ldots u^2_{j_n}\;.
\end{equation}
Now, the harmonic derivative converts any $u^2$ into $u^1$, so eq.
(\ref{C1}) implies the vanishing of all the terms in
(\ref{infexp}) but the first one,
\begin{equation}\label{shexp}
  D^{\,1}_2\phi^1(x,u) = 0 \ \Rightarrow \ \phi^1(x,u) =
\phi^i(x)u^1_i\;.
\end{equation}
The same argument can be applied to the higher-order components of
the superfield. Thus, the spinor component $\psi_\alpha(x,u)=
D^2_\alpha q^1\vert_{\theta=0}$ satisfies the constraint
\begin{equation}\label{spcons}
  D^{\,1}_2\psi_\alpha(x,u) = D^{\,1}_2 D^2_\alpha q^1\vert_{\theta=0} =
D^1_\alpha q^1\vert_{\theta=0} = 0
\end{equation}
(see (\ref{C0})). Since this component is chargeless, the harmonic
condition (\ref{spcons}) implies that it is a singlet,
$\psi_\alpha(x,u)=\psi_\alpha(x)$. In a similar way one can find
the complete expansion of the on-shell superfield:
\be
q^1=\varphi^i(x)u^1_i +\tta\psi_\alpha(x)+\btoa\bar{\chi}_\da(x)
+i\tta\btob\padb\varphi^i(x)u^2_i \lb{C5} \ee where all the
components satisfy their free massless equations of motion. We
clearly see that this superfield is ``short" in the sense that it
only depends on half of the Grassmann variables of the full $N=2$
superspace. It is even ``ultrashort" in the sense that the top
spin in it is $1/2$ instead of the maximal spin 1 allowed in a
G--analytic scalar superfield.

It is useful to note the relations
\begin{equation}\label{C2}
  (D^2)^2 q^1=(\bar{D}_1)^2 q^1=0
\end{equation}
which can again be derived from the basic constraints. They are
the covariant form of the statement that the superfield is linear
in $\theta_2$ and $\bto$, as can be seen from the expansion
(\ref{C5}).

When rewritten in the central basis coordinates
$x,\theta_i,\bar\theta^i$,  the on-shell hypermultiplet superfield
recovers its original form
\be
q^1=u^1_i q^i(z)\lb{C5b} \ee of an $SU(2)$ doublet. In fact, this
observation can be given an invariant meaning as follows. We
remark that the harmonic derivatives $D^{\,I}_J$ form the algebra
of $SU(2)$
\begin{equation}\label{algsu2}
  [D^{\,I}_J,\;D^{\,K}_L] = \delta^K_J D^{\,I}_L -
\delta^I_L D^{\,K}_J
\end{equation}
realized on the superfield $q^1$. As explained in the Appendix,
the derivative $D^{\,1}_2$ is the positive root (``creation
operator") of this algebra. Then the condition (\ref{C1}) simply
defines the highest weight of an $SU(2)$ representation. The
quantum number associated to this representation (``superisospin")
coincides with this of the first component. To see this we write
down the Casimir of this $SU(2)$,
\begin{equation}\label{cassu2}
  C_2 = D^{\,I}_J D^{\,J}_I = {1\over 2}(D^0)^2 + D^0 + 2
D^{\,2}_1 D^{\,1}_2
\end{equation}
where $D^0 = D^{\,1}_1 - D^{\,2}_2$ is the $U(1)$ charge operator.
Then, applying this Casimir to the on-shell superfield $q^1$
satisfying the H-analyticity constraint (\ref{C1}) and using the
fact that it carries a definite $U(1)$ charge, $D^0 q^1 = q^1$, we
obtain
\begin{equation}\label{caseig}
  C_2 q^1 = {3\over 2}q^1\;.
\end{equation}
We see that $q^1$ is an eigenfunction of the Casimir, realizing
the doublet (isospin 1/2) representation, just like its first
component (see (\ref{shexp})). The important point here is that
the harmonic derivatives $ D^{\,I}_J$ commute with the
supersymmetry generators, therefore $C_2$ is the  superisospin
Casimir of the entire Poincar\'{e} supersymmetry algebra. Note that
this algebra has another Casimir, that of superspin. We can apply
similar arguments to compute the value of this Casimir on the
superfield $q^1$. Indeed, the G-analyticity conditions (\ref{C0})
are equivalent to demanding that the positive odd roots of the
Poincar\'{e} supersymmetry algebra annihilate  $q^1$. This ensures
that the superspin also takes a definite eigenvalue, so the
on-shell superfield $q^1$ realizes an irrep of Poincar\'{e}
supersymmetry (see in this context Ref. \cite{RiSo} for a general
discussion of irreducibility conditions on superfields).

The above analysis can be repeated for other superfields
satisfying both the G-- and H--analyticity conditions but carrying
different $U(1)$ charges. Take, for instance, the off-shell linear
multiplet. In HSS it is described by a G--analytic superfield
$L^{11}$ of $U(1)$ charge $+2$  \cite{GIOS}. Unlike the
hypermultiplet $q^1$, this time the H--analyticity condition
\be
\Dot L^{11}=0\lb{C9} \ee does not put the superfield  on shell.
The only restriction on the components involving space-time
derivatives is that the vector in  $L^{11}$ must be
divergenceless. The $SU(2)$ Casimir now takes the eigenvalue $4$
which corresponds to isospin $1$ (triplet irrep). To put this
superfield on shell, an additional constraint is required,
\be
(D^2)^2  L^{11}=0\lb{C10}~. \ee This H--analytic superfields of
charge $+2$ is not ultrashort. H--analyticity for superfields of
charge $q\geq +3$ still makes them irreducible but does not yield
any constraints on the remaining components.

Now we turn to the other basic multiplet of $N=2$ supersymmetry,
that of SYM. The free on-shell ultrashort Maxwell multiplet is
described by a chiral (harmonic independent) superfield satisfying
an additional second-order constraint
\be
\bar{D}_{\da k}W=0~,\q D^{\alpha(i}D^{j)}_\alpha W=0~.\lb{C6} \ee
In the chiral basis, the on-shell components of this superfield
are \be W = \phi(x) + \theta^\alpha_i\lambda_\alpha^i(x) +
\theta^{(\alpha i}\theta^{\beta)}_i F_{\alpha\beta}~.\lb{C7} \ee
This is another example of an ultrashort multiplet (its expansion
goes only up to $\theta^2$, as compared to $\theta^4$ for a
generic chiral $N=2$ superfield).

We do not consider here the non-Abelian generalization of the HSS
description of the hypermultiplet and gauge multiplet
\cite{GIK1,Z2}. It should be stressed that the linear
H--analyticity condition \p{C1} is valid for gauge-invariant
superfields only, otherwise it should contain a harmonic
connection.

We can now use the two objects above, the on-shell hypermultiplet
$q^1$ and the SYM field strength $W$ as building blocks which will
allow us to construct all the short representations of
$SU(2,2/2)$. By adapting the series A), B), C) from Section 2 for
$N=2$ we have \cite{FZ}:
\begin{eqnarray}
&\mbox{A)}& \ell\geq 2+2J_2-r+2I \geq 2+2J_1+r+ 2I, \quad
J_1J_2\geq 0\;; \nonumber\\ &\mbox{B)}& \ell= -r+2I \geq 2+2J+r+2I
, \quad J_1=J,\, J_2= 0\;; \nonumber\\ &\mbox{C)}& \ell = 2I ,
\quad J_1= J_2=r= 0\;.
\end{eqnarray}
A general long multiplet, belonging to the A) series, contains 4
$\theta$'s, 4 $\bar\theta$'s and $J_{max}=2$ (in the case of
$J_1=J_2=0$ for the highest weight state, {\ie}   for the $\theta
= \bar\theta =0$ superfield component). In terms of our building
blocks this series corresponds to chiral-antichiral multiplication
of the type $W\bar W$, or to analytic-antianalytic multiplication
of hypermultiplets of the type $q^1{q^2}$ (where $q^2 =
D^{\,2}_1q^1$ is a superfield satisfying ``antianalyticity"
constraints),  or to products of both.

Series B) for $J=0$, $I =0$  corresponds to chiral multiplets
which can be obtained by the following operator series:
\be
\trace [W^p] \ee ($p=1$, $\ell =-r=1$ is the on-shell SYM multiplet
itself defined in \eq{C6}). These superfields depend on the 4
left-handed $\theta^\alpha_i$ and one immediately sees that the
top spin in their expansion belongs to the Lorentz representation
$(1,0)$. The above series is chiral (short) and for $N=2$ may be
called ``tensor multiplet'' tower \cite{grw}, since the maximum
spin is $(1,0)$ (with $I=0$ and $\ell_{(1,0)}=1+p$).

Series C) corresponds to the analytic multiplication of
hypermultiplets:
\be
\mbox{Inv} (q^1)^{2I}  \label{hyper} \ee where the symbol Inv
means a gauge invariant product. The case $I={1/2}$ corresponds to
the on-shell hypermultiplet itself. The lowest component of these
superfields is in the isospin $I$ $SU(2)$ representation. The
superfields depend on the 2 left-handed $\theta^\alpha_2$ and the
2 right-handed $\bar\theta^{1\dot\alpha}$, so the top spin (for $I
> \o 12$) is a $(\o 12 , \o 12 )$ vector in the isospin $I -1$
$SU(2)$ representation with dimension $\ell_{(\o 12 , \o 12 )} =2I
+1$. For $I =1$ the vector is ``conserved'' and gives a ``current"
superfield (this is the linear multiplet (\ref{C9})).

There is another intermediate shortening in the B) series ($J=0$)
obtained by multiplying chiral with G--analytic superfields:
\be
\mbox{Inv} [W^p (q^1)^{2I}]  \, .\label{intermediate} \ee This
superfield is G--analytic in a weaker sense than either $W$ or
$q^1$, satisfying only the constraint (dropping the Inv symbol)
\begin{equation}\label{intermediate'}
  \bar
D_{2\dot\alpha} [W^p (q^1)^{2I}]=0\;,
\end{equation}
and so it depends on the 4 $\theta^\alpha_{1,2}$ and on the 2
$\bar\theta^{1\dot\alpha}$. Thus, the top spin in it is $\o 32 =
(1,\o 12)$.

There are other even shorter multiplets when the
component fields satisfy ``space--time constraints'' {\ie}
conservation laws (transversality) or equations of motion. This
happens when the dimension takes a particular value.

In A) $W\bar W$ has $\ell =2$ ($J_1=J_2=0$), $r=0$, $I =0$ and corresponds to
the conserved stress-tensor multiplet.
There is a similar object in the analytic-antianalytic multiplication of
$q^1$ and $q^2=\Dto q^1$ .

The superfield $W^pq^1$ satisfies the additional linearity
constraint (see (\ref{C2}) and (\ref{C6}))
\be
 (\bar{D}_1)^2 [W^pq^1]=0~, \ee so it
depends on the 4 $\theta$'s and only linearly on $\bar\theta_1$.
In B) $\ell =1$ and in C) $I =\o 12$ correspond to the basic
``super-singleton'' UIR of $SU(2,2/2)$. In the $AdS/CFT$ language
the $\ell =2, r=J_1=J_2=I =0$ (in A)) and $\ell =2, r=J_1=J_2 =0,
I =1$ (in C)) correspond to massless graviton and gauge bosons
(``current superfields''). Semishort multiplets, obeying to ``A)
treshold'', also exist. They are $\bar W W ^n $, $n>1$, ($\bar
D^{(i}\bar D^{j)}(\bar W W^n)=0$) with $\ell =1+n$, $r=1-n$ {\ie}
$\ell =2-r$.


\subsection{The $N=3$ case}
\subsubsection{The $N=3$ super Yang--Mills multiplet}
The $N=3$ Yang--Mills multiplet is described by the field strength
superfield \cite{HST,GIO}
 $W_{ij}(x,\theta)\equiv \varepsilon_{ijk} W^k$ defined
by anticommuting the gauge-covariant spinor derivatives:
\begin{equation}
  \{\bar{\nabla}_{i\da},\bar{\nabla}_{j\db}\} =
\epsilon_{\da\db}W_{ij}\;
\end{equation}
or by the conjugate superfield $\bar{W}^{ij}$. It the Abelian case
this superfield  satisfies the following on-shell constraints:
\bea &&D_\alpha^i W_{jl}={1\over2}(\delta^i_j D_\alpha^k W_{kl}-
\delta^i_l D_\alpha^k W_{kj}) ,\lb{3a'} \\ &&\bar{D}_{i\da}
W_{jk}+\bar{D}_{j\da} W_{ik}=0 \lb{3b} \eea and their complex
conjugates.

The $SU(3)/U(1)\times U(1)$ HSS has been introduced for the
off-shell description of the $N=3$ Yang-Mills theory \cite{GIO}.
Here we shall use this superspace for the classification of the
short on-shell $N=3$ multiplets. Some basic facts about $N=3$ HSS
are given in the Appendix.

One can define three different harmonic projections of the Abelian
on-shell superfield $W_{ij}$:
 \bea
 && W_{23}\equiv W^1 =u_2^i u_3^j W_{ij}~,\lb{d1}\\
&& W_{13}\equiv -W^2=u_1^i u_3^j W_{ij}~,\lb{d3}\\ &&
W_{12}=W^3=u_1^i u_2^j W_{ij}~.\lb{d2}
 \eea
By projecting the on-shell constraints (\ref{3a'}), (\ref{3b}) 
with the appropriate harmonics one finds sets of G--analyticity 
constraints on each of these superfields. They lie in three 
different analytic superspaces with six odd coordinates (see the 
Appendix). The existence of such analytic superspace involving 
unequal numbers of left- and right-handed odd variables was first 
pointed out in \cite{GIO1} and then generalized to the so-called 
$(N,p,q)$ superspaces in \cite{hh}. 

Consider, for example, the superfield $W^1$ satisfying the 
following conditions of G--analyticity \cite{hh} 
\be
\bDta W^1=\bDha W^1=\Doa W^1=0\lb{d4} \ee meaning that
\begin{equation}\label{d41}
 W^1 = W^1(x_\A,\theta_2,\theta_3,\bar\theta^1,u)
\end{equation}
in the appropriate analytic basis (\ref{a14}).

The G--analytic superfield $W^1$ is a harmonic superfield with an
infinite expansion on the harmonic coset. In order to get back the
original constrained harmonic-independent superfield
$W_{ij}(x,\theta)$ we need to impose conditions of H--analyticity.
To this end we should use only the harmonic derivatives
corresponding to the positive roots of $SU(3)$ (see the Appendix).
These are:
\be
\Dot W^1=\Dth W^1=\Doh W^1=0~. \lb{d5} \ee As expected, they form
a closed algebra (CR structure) with the spinor derivatives in
(\ref{d4}), {\ie} preserve G--analyticity. Note that only the
first  of eqs. (\ref{d5}) is the true equation of motion. The
second one is purely kinematical and the third one is a corollary
of the first two, since $\Doh =[\Dot,\Dth]$.

The H--analyticity conditions (\ref{d5}) have the meaning of
$SU(3)$ irreducibility conditions. Indeed, the derivatives
$D^{\,I}_J$ form the algebra of $SU(3)$
\begin{equation}\label{algsu2'}
  [D^{\,I}_J,\;D^{\,K}_L] = \delta^K_J D^{\,I}_L -
\delta^I_L D^{\,K}_J
\end{equation}
realized on the superfield $W^1$. Then (\ref{d5}) just defines the
highest weight of an irrep. To find out which one, we can write
down the Casimirs of this $SU(3)$,
\begin{equation}\label{d101}
  C_2 = D^{\,I}_{J} D^{\,J}_{I}\;, \qquad
C_3 = D^{\,I}_{J} D^{\,J}_{K}D^{\,K}_{I}
\end{equation}
and rearrange the derivatives so that all the analytic ones from
eq. (\ref{d5}) are on the right. Then, applying these Casimirs to
the on-shell superfield $W^1$ and using the fact that it carries a
definite $U(1)\times U(1)$ charge,
\begin{equation}\label{u1ch}
 D^{\,1}_{1} W^1 = W^1\;, \quad
D^{\,2}_{2} W^1=D^{\,3}_{3} W^1=0\;,
\end{equation}
we find that  $W^1$ is an eigenfunction of the Casimirs. Since the
harmonic derivatives are supersymmetric invariant, we can switch
back to the basis in superspace where the $\theta$'s are  not
projected with harmonics. There $W^1 = W^iu^1_i=u_2^i u_3^j
W_{ij}$ and we come back to the original form (\ref{d1}). Thus,
the super-$SU(3)$ quantum numbers of the superfield $W^1$ coincide
with those of its first component.

The G--analytic superfield is also an eigenfunction of the
superspin Casimir. The reason is that it is annihilated by half of
the odd generators (spinor derivatives), so it is a highest weight
of the entire $N=3$ Poincar\'{e} supersymmetry algebra. Moreover, the
close examination of the components below shows that they are on
shell, satisfying massless equations of motion. Thus, $W^1$
realizes an irrep of conformal supersymmetry as well.

It is easy to prove that $W^1$ also obeys linearity conditions
with respect to each $\theta$:
\be
(D^2)^2 W^1=(D^2D^3) W^1=(D^3)^2 W^1=(\bar{D}_1)^2 W^1=0\;.\lb{d6}
\ee This is done by using the harmonic derivatives, e.g.,
\begin{equation}\label{d8}
  \Dot (D^2)^2 W^1=2(D^1D^2) W^1
=0~\Rightarrow~(D^2)^2 W^1 =0~.
\end{equation}

Further, by examining the components of the superfield $W^1$ one
finds that the top chargeless component lies very low in the
$\theta$ expansion:
\be
F_{\alpha\beta}^+ = D^2_{(\alpha}D^3_{\beta)}W^1\vert_0~.\lb{d10}
\ee
Acting with harmonic derivatives on the higher-order spinor
derivatives (components) of $W^1$ one can easily show that all of
them are expressed in terms of space-time derivatives of the
preceding components. In this way, one finally obtains the
components of the ultrashort on-shell superfield
$W^1$:
\begin{eqnarray}
&&W^1 = \phi^1 + \btoa\bar{\lambda}_\da + \tta \lambda_{3\alpha} -
\tha \lambda_{2\alpha}
  -i \tta\btob\padb  \phi^2
 \nonumber\\
  &&+
\theta_2^\alpha\theta_3^\beta F_{\alpha\beta}^+ -i
\bar\theta^{1\dot\alpha}\theta^\alpha_2\theta_3^\beta
\partial_{(\alpha\dot\alpha}\lambda_{1\beta)}\;.
  \lb{d10b}
\end{eqnarray}
where the physical fields satisfy massless field equations.

One can treat the projection $W^2=-W_{13}$ in a similar way. The
spinor and harmonic  derivatives annihilating $W^2$ are
\be
\Dta,~\bDoa,~\bDha,~\Doh,~\Dto,~\Dth~.\lb{d11} \ee The harmonic
conditions make the leading  component $[W^2]|_0$ an irrep of
$SU(3)$, and thus give definite super-$SU(3)$ quantum numbers to
the whole superfield. The corresponding linearity conditions are
\be
(D^1)^2 W^2=(D^1D^3) W^2=(D^3)^2 W^2=(\bar{D}_2)^2
W^2=0\;.\lb{d12} \ee The Abelian superfield $W^2$ lives in a
rotated version of the G--analytic superspace (\ref{a14}):
 \bea &&
W^2(x^\prime_\A,\theta_1,~\theta_3,~\bar{\theta}^2,~u)\\ &&
x^\prime_\A=x+i\theta_1\bto-i\theta_2\btt+i\theta_3\bth~. \eea The
components of $W^2$ are obtained from those of $W^1$ (\ref{d10b})
by exchanging 1 with 2.
It is evident that $W^2=\Dto W^1$, so both superfields describe
the same on-shell vector multiplet.

Finally, consider the harmonic projection of the $N=3$ superfield
$\bar{W}^{ij}$
\be
\bar{W}^{12}=u^1_i u^2_j\bar{W}^{ij}~.\lb{d13} \ee The on-shell
constraints on $\bar{W}^{ij}$ are equivalent to the following
G--analyticity conditions:
\be
\Doa\bar{W}^{12}=\Dta\bar{W}^{12}=\bDha\bar{W}^{12}=0~.\lb{d13'}
\ee This  Abelian superfield lives in yet another version of the
G--analytic superspace (\ref{a14}),
\be
\bar{W}^{12}(x''_\A,\theta_3,\bto ,\btt ,u)\lb{d14}\;. \ee In
addition, it satisfies the harmonic constraints \be
 D_3^2\bar{W}^{12}
=D^1_2\bar{W}^{12}=D^1_3\bar{W}^{12} = 0\;.\lb{d15} \ee From these
constraints follow the linearity conditions
\be
(D^3)^2\bar{W}^{12} =(\bar{D}_1)^2 \bar{W}^{12}
=(\bar{D}_2)^2\bar{W}^{12} =0~.\lb{d16} \ee This superfield has
the following components:
\begin{eqnarray}
&&\bar{W}^{12} = \bar\phi^{12} - \tha\lambda_\alpha - \btoa
\bar\lambda^2_\da  + \btta \bar\lambda^1_\da  +i \tha\btob\padb
\bar\phi^{23}    -i\tha\bttb \padb  \bar\phi^{13} \nonumber\\
&&-\bar\theta^{1\dot\alpha}\bar\theta^{2\dot\beta}
F^-_{\dot\alpha\dot\beta} +i
\bar\theta^1_{\dot\alpha}\bar\theta^2_{\dot\beta}\theta^\alpha_3
\partial_\alpha^{(\dot\alpha}\bar\lambda^{3\dot\beta)}\;.
  \lb{d17}
\end{eqnarray}
Once again, this is another equivalent description of the same
$N=3$ on-shell SYM multiplet.

\subsubsection{Series of short $N=3$ multiplets}

The A), B), C) UIR's of the $SU(2,2/3)$ algebra are given by
adapting the quantities $X,Y$ to the $N=3$ case with $m_1 =a+b$,
$m_2 =a$ where $[a,b]$ are the $SU(3)$ Dynkin labels.
\\
It then follows that ($m=m_1+m_2=2a+b$):
\begin{itemize}
\item{
\be
\mbox{A)}\ \ \ell \geq 2+2J_2-r+ \o 43 a +\o 23 b  \geq
  2+2J_1+r + \o 23 a +\o 43 b
\label{3x} \ee which implies:
\begin{eqnarray}
-r&\geq & J_1 -J_2 -\o 13(a-b) \nonumber\\ \ell &\geq &2+  J_1 +J_2
+a+b
\end{eqnarray}
(or $r\to -r$, $J_1 \to J_2$, $a\to b$)}
\item{
\begin{eqnarray}
\mbox{B)}&& J_1 =J \, , J_2=0 \nonumber\\ && \ell = -r + \o
43 a +\o 23 b \geq 2+2J+r + \o 23 a +\o 43 b
\label{bseries}
\end{eqnarray}
which implies:
\begin{eqnarray}
-r&\geq & 1+J-\o 13(a-b) \nonumber\\ \ell &\geq & 1+J+a+b
\end{eqnarray}
(or $J_1\to J_2 $, $r\to -r$. $a\to b$) }
\item{
\be
\mbox{C)} \quad J_1=J_2=0 , \quad \ell =a+b , \quad r=\o 13(a-b)
\label{c3} \ee The Yang--Mills (supersingleton)
 multiplet corresponds to series C)
for $a=0$, $b=1$, $r=-\o 13$.}
\end{itemize}

Now, let us realize these abstract short representations in terms
of the SYM superfields $W$. The series C) for $a=0$ corresponds to
the tower with maximal shortening:
\be
\trace (W^1)^b= C[0,b]\;. \label{3c} \ee  This is a superfield
depending on 4 $\theta$'s and 2 $\bar\theta$'s, consequently the
maximum spin ($b>1$) is $J=\o 32$ in the Lorentz representation
($1,\o 12$). The first component is a scalar with dimension
$\ell=b$ and $r$-charge $r=-b/3$ in the $[0,b]$ UIR of $SU(3)$.
The case $b=1$ is the ultrashort Yang--Mills singleton $W^1$.

The short H--analytic superfields  for $b=2, 3$   have the the
following chargeless components:
\begin{equation}\label{d19}
  (D^2)^2(D^3)^2 C[0,2]\lb{d19'}~, \quad
D^2_{(\alpha}D^3_{\beta)}(\bar{D}_1)^2 C[0,3]~.
\end{equation}
All the higher components are space-time derivatives of the lower
ones. In the case $b\geq 4$ H--analyticity does not lead to any
extrashortening, e.g. the superfield $C[0,4]$ contains an
independent top component
\be
(D^2)^2(D^3)^2(\bar{D}_1)^2 C[0,4]~.
\ee

The complete C) series of short multiplets can be obtained by
taking the products
\be
(W^1)^b  (\bar W^{12})^a = C[a,b]
 \label{3semilong}
\ee (we omit the traces).
The first components of these
superfields contain {\it analytic} harmonics with $a+b$ indices ) 1
and $a$ indices 2 corresponding to the UIR $[a,b]$ of $SU(3)$.
We obtain the generic short operator of the C) series
with $\ell =a+b$, $r=\o 13(a-b)$ and with $J_{max}=2=(1,1)$, since
this operator contains 4 $\theta$, 4 $\bar\theta$.

Now, let us consider $C[a,b]$ as an abstract G--analytic
superfield
$$
C[a,b]=C[a,b](\theta_2,\theta_3,\bar\theta^1,\bar\theta^2)
$$ with the given $SU(3)$ quantum numbers. The H--analyticity
conditions for these representations are the same as those for the
building block $W^1$ (see (\ref{d5})),
\be
\Dot  C[a,b]= \Dth C[a,b] = D^1_3 C[a,b] =0~.\lb{d23} \ee This is
equivalent to imposing an $SU(3)$ irreducibility condition. Using
both G-- and H--analyticity one can derive various constraints on
the components. For instance,  we  find the following highest
chargeless components in the simplest cases: \bea
&&D^2_{(\alpha}D^3_{\beta)}\bar{D}_{1(\da}\bar{D}_{2\db)}C[1,1]
~,\lb{d24}\\ &&(D^2)^2(D^3)^2\bar{D}_{1(\da}\bar{D}_{2\db)}C[1,2]
~,\lb{d25} \\ &&(D^2)^2(D^3)^2(\bar{D}_1)^2\bDta C[1,3] ~.\lb{d26}
\eea {}From these levels of the expansion on the corresponding
superfields should become short. Note, however, that there is an
additional linearity constraint in the case $a=1$,
\be
(\bar{D}_2)^2 C[1,b]=0~, \ee which follows from the properties of
$\bar W^{12}$ in the product \p{3semilong} but cannot be obtained
from H--analyticity alone.

The multiplet which is dual to the graviton multiplet of $N=6$
supergravity in $AdS_5$ is given in \eq{d24} ($C[1,1]$). Indeed
the top component is the spin 2 $(1,1)$ graviton multiplet with
$\ell =4$.
This is a ``current superfield''.

Note that \eq{3semilong} for $b=a$ is invariant under an additional $r$-phase
$W_i \to e^{\ii\alpha}
W_i$ which commutes with the $SU(2,2/3)$ algebra \cite{fpz,intr}.

By selecting the $r$ invariant singlets $C[a,a]$ we obtain a
tower of spin 2 short multiplets which are $\o 13$ BPS states of
$N=6$ supergravity \cite{fpz,HST}. Other cases in this class of
analytic representations have no semishortening.

The next series of representations is given by the products
\be
(W^1)^{a+b}(W^2)^a = B[a,b]~.\lb{d29} \ee This superfield 
satisfies only one G--analyticity condition, 
\begin{equation}\label{onlyone}
  \bDha B[a,b]=0 \ \Rightarrow \  B[a,b]=
B[a,b](\theta_1,\theta_2,\theta_3,\bar\theta^1,\bar\theta^2)\;.
\end{equation}
This implies that the top spin in it is $J_{max}= {5\over 2}=
({3\over 2},1)$. Further, the same H--analyticity constraints as
in  (\ref{d5}),
\begin{equation}\label{follhan}
  \Dot B[a,b] = \Dth B[a,b] = \Doh B[a,b] = 0
\end{equation}
imply that the first component belongs to the UIR $[a,b]$. The
dimension and $r$-charge of this superfield are
\be
\ell =2a+b~,\qq r=-{1\over3}(2a+b)~.\lb{Bdim} \ee

One can prove the following constraints: \bea &&
\bar{D}_{1(\da}\bar{D}_{2\db)}B[a,b]=0~,\\ &&(\bar{D}_1)^2\bDta
B[a,b]=(\bar{D}_2)^2\bDoa B[a,b]=0~.\lb{3rest} \eea Thus, this
representation can contain all 6 $\theta$'s and bilinear scalar
combination  $(\bar{\theta}^1\bar{\theta}^2)$.

The superfields $B[1,b]$  satisfy the following additional
conditions:
\be
(D^1)^2 B[1,b]=(\bar{D}_2)^2 B[1,b]=0 \ee which follow from the
properties of $W^2$. The superfield $B[1,0] = W^1W^2$ is even
further constrained:
\be
(D^2)^2 B[1,0] =(\bar{D}_1)^2 B[1,0] =0~. \ee

We find the following highest chargeless components: \bea
&&D^1_{(\alpha}D^3_{\beta)}\bDoa  B[1,0]~,\lb{d31}\\
&&D^1_\alpha D^2_\beta(D^3)^2 \bDoa  B[1,1]~,\lb{d32}\\
&&(D^1D^2)(D^3)^2 (\bar{D}_1)^2 B[2,1]~.\lb{d33} \eea This means
that these superfields are semishort.

These representations belong to the B) series ($J=0$).

The series A) corresponds typically to superfields with 6
$\theta$, 6 $\bar\theta$ and for $J_1=J_2=0$ contains multiplets
with $J_{max}=3=(\o 32, \o 32)$.

The last series (still corresponding to the B) shortening)
 can be constructed by taking the products 
\be
(W^1)^{p+q+n}(W^2)^{q+n}(\bar{W}^{12})^{n} = B'[q+2n,p]~, \q n\geq 
1~. \lb{last} \ee It lives in the same superspace with 10 spinor 
coordinates as $B[a,b]$ and has its first component in the $SU(3)$ 
UIR $[q+2n,p]$. It is clear that one can find members of both 
series having their first components in the same UIR, $B'[q+2n,p]$ 
and $B[q+2n,p]$. However, the two short multiplets are not 
equivalent. The dimension and $r$-charge of the $B'[q+2n,p]$ 
series are $\ell =p+2q+3n$, $r=-{1\over3}(p+2q+n)$ whereas for the 
$B[q+2n,p]$ series they are $\ell =p+2q+4n$, 
$r=-{1\over3}(p+2q+4n)$.

In conclusion we can say that the short analytic $N=3$
representations are defined by the choice of the lowest harmonic
representation, the Grassmann dimension and the quantum numbers
$\ell$ and $r$.

\subsection{The $N=4$ case}

\subsubsection{The $N=4$ SYM multiplet}

The $N=4$ Yang--Mills multiplet is described by the field strength
superfield $W^{ij}(x,\theta)$
satisfying the reality condition \cite{So,HST}
$$W^{ij}= {1\over2}\epsilon^{ijkl}
W_{kl}~,\qq W_{kl}=\bar{W^{kl}}
$$
and the following on-shell constraints: \bea &&\bar{D}_{i\da}
W^{jk}={1\over 3}(\delta^j_i \bar{D}_{l\da} W^{lk}- \delta^k_i
\bar{D}_{l\da} W^{lj})~ , \lb{e1}\\ &&D^i_\alpha W^{jk}+D^j_\alpha
W^{ik}=0~.\lb{e2} \eea Note that both forms of $W$ contain
$F_{\alpha\beta}$ and $F_{\da\db}$, so we do not use
$\bar{W}$ for $N=4$ superfields.

We shall rewrite these constraints in $N=4$ HSS. To this end we 
have to chose one of the harmonic coset spaces for the group 
$SU(4)$ listed in Ref. \cite{IKNO}. It should be pointed out that 
a harmonic interpretation of the $N=4$ SYM constraints has for the 
first time been proposed in Refs. \cite{hh}. It makes use of the 
harmonic coset $SU(4)/S(U(2)\times U(2))$. This is sufficient to 
show that the on-shell $W$ is a G--analytic superfield depending 
only on half of the odd variables. However, the residual symmetry 
$S(U(2)\times U(2))$ in the approach of Ref. \cite{hh} turns out 
too restrictive for the analysis of all representations. In order 
to have maximal flexibility we shall use harmonics on the coset 
$SU(4)/ U(1)^3$ (see the Appendix for details of the definition 
and the basic properties). 

With the help of these harmonics we can introduce three
independent projections of the on-shell field strength:
\begin{eqnarray}
W^{12} &\equiv & u^1_i u^2_j W^{ij}=-W_{34}
\lb{e3} \\ W^{13} &\equiv & u^1_i u^3_j W^{ij}=W_{42}\lb{e4}  \\
W^{23} &\equiv & u^2_i u^3_j W^{ij}=W_{14}\lb{e5}~.
\end{eqnarray}
It is easy to see that  the constraints on  $W^{ij}$ imply that
these three superfields belong to three different G--analytic
subspaces of HSS. For example, the projection $W^{12}$ satisfies
the  G--analyticity constraints corresponding to the following
spinor derivatives:
\be
\Doa W^{12}=\Dta W^{12} =\bDha W^{12} =\bDfa W^{12} =0\lb{e6}~.
\ee In the appropriate basis in superspace (\ref{a14}) the
analytic $W^{12}$ has the form
\be
W^{12} =
W^{12}(x_\A,\theta_3,\theta_4,\bar\theta^1,\bar\theta^2,u)~.\lb{e8}
\ee
We see that $W^{12}$ depends on only 8 out of the 16
$\theta$'s of the full $N=4$ superspace. It is then obvious that
its $\theta$  expansion can in principle go up to spin 2:
\begin{equation}\label{e63}
 W^{12} = \ldots + \theta^\alpha_3\theta^\beta_4
\bar\theta^1_{\dot\alpha}  \bar\theta^2_{\dot\beta}
A^{\dot\alpha\dot\beta}_{\alpha\beta} + \ldots\ .
\end{equation}
This is an example of a short multiplet (a generic $N=4$
superfield expansion goes up to spin 4). In fact, $W^{12}$ is even
shorter, as we shall see in the next subsection.

In order to achieve equivalence with the original constraints
(\ref{e1}), (\ref{e2}) we have to eliminate the non-trivial
harmonic dependence of $W$. This is done by imposing conditions of
H--analyticity, in addition to G--analyticity. As in the cases
$N=2,3$, we choose the set of six harmonic derivatives
corresponding to the positive roots of $SU(4)$:
\begin{equation}\label{e61}
  D_J^{\, I}W^{12}=0\;, \qquad I,J=1,2,3,4, \ \ I<J\;.
\end{equation}
They define the highest weight of an $SU(4)$ irrep. In fact, among
them only three are independent, $(D_2^{\, 1}\;,\ D_3^{\, 2}\;,\
D_4^{\, 3})W^{12}=0$, but it is often convenient to use all the
six. The implications of the condition $D_2^{\, 1}W^{12}=0$ on the
leading component in the $\theta$ expansion of
$W^{12}=\phi^{12}(x,u)+\ldots\ $ are easy to see:
\begin{equation}\label{e62}
  D_2^{\, 1}(\phi^{ij}(x)u^1_iu^2_j) =
\phi^{ij}(x)u^1_iu^1_j = 0\ \Rightarrow \phi^{ij} = -\phi^{ji}
\end{equation}
since the harmonic variables commute. In other words, the
component $\phi$ is in the $SU(4)$  UIR $[0,1,0]$. The remaining
harmonic conditions eliminate any dependence on the other
harmonics in $\phi^{12}(x,u)$. The same argument shows that the
remaining components of the superfield either belongs to UIR's of
$SU(4)$(if they are not expressed in terms of the lower components
or just vanish), so that in the end the entire superfield recovers
its original trivial harmonic dependence shown in eq. (\ref{e5}).

The harmonic conditions (\ref{e61}) ensure that the superfield
$W^{12}$ forms a representation of supersymmetry with fixed
$SU(4)$ super-quantum numbers. Indeed, the harmonic derivatives
$D^{\,I}_J$ form the algebra of $SU(4)$,
\begin{equation}\label{F888}
  [D^{\,I}_J,\;D^{\,K}_L] = \delta^K_J D^{\,I}_L -
\delta^I_L D^{\,K}_J
\end{equation}
realized on $W^{12}$. At the same time, these derivatives are
super-covariant, {\ie} commute with the supersymmetry generators.
Therefore the $SU(4)$ Casimir operators
\begin{equation}\label{e101}
  C_n = D^{\,I_1}_{I_2} D^{\,I_2}_{I_3}\ldots D^{\,I_n}_{I_1}\;,
\qquad n=2,3,4
\end{equation}
are automatically super Casimirs. Now, the $SU(4)$ algebra
(\ref{F888}) allows us to rewrite (\ref{e101}) in such a way that
all the $D^{\,I}_J$ with $I<J$ appear on the right, after which we
can make use of the conditions (\ref{e61}). Thus, the Casimirs are
reduced to polynomials of the charge operators $D^{\,I}_I$ and
take eigenvalues on the superfield  $W^{12}$ determined by its
charges. The conclusion is that the supermultiplet described by
$W^{12}$ has definite $SU(4)$ quantum numbers which coincide with
those of its first component.

In exactly the same way one can show that the other two
projections of the field strength live in the two alternative
G--analytic subspaces involving only 8 $\theta$'s each: \bea &&
\label{e71} W^{13}(x'_\A,\theta_2,\theta_4,\bar\theta^1,
\bar\theta^3,u)~,\\
&&x'_\A=x-i(\theta_1\bar\theta^1+\theta_3\bar\theta^3-\theta_2\bar\theta^2
-\theta_4\bar\theta^4)~;\nn\\ && \label{e72}
W^{23}(x''_\A,\theta_1,\theta_4,\bar\theta^2, \bar\theta^3,u)~,\\
&&x''_\A=x-i(\theta_1\bar\theta^1+\theta_3\bar\theta^3-
\theta_2\bar\theta^2 -\theta_4\bar\theta^4)~,\nn \eea where the
corresponding G--analytic bases in superspace have been used.

In addition, these G--analytic superfields satisfy H--analyticity
conditions which can be obtained from eq. (\ref{e61}) by permuting
the indices. As before, they make the superfield an irrep of
$SU(4)$. It should be stressed that these analyticity conditions
are flat (linear) for all $W$'s in the Abelian theory or when
applied to gauge-invariant composite operators of the type
$\mbox{Tr}W^n$ in the non-Abelian theory.

\subsubsection{Series of short multiplets}

The UIR's of the $PSU(2,2/4)$ superalgebra fall in three classes
\cite{dp,FZ}:
\begin{itemize}
\item{
\be
\mbox{A)} \quad \ell \geq 2+J_1+J_2+a+b+c \, , \quad J_2-J_1 \geq
\o 12 (c-a) \quad (\mbox{or } J_1\leftrightarrow J_2 , a \leftrightarrow c)
\ee
Massless bulk multiplets correspond to maximal shortenings with $J_2=J_1$,
$a=b=c=0$}
\item{
\be
\mbox{B)} \quad \ell =\o 12 (c+2b+3a) \geq 2+2J +\o 12 (3c+2b+a) \, ,
\quad J_1=J \, , J_2=0
\label{b4}
\ee
(or $J_1\leftrightarrow J_2 $, $ a \leftrightarrow c$)
with $\ell \geq 1+J+a+b+c $, $1+J \leq \o 12(a-c)$ }
\item{
\be
\mbox{C)} \quad \ell =2a+b  \, , \quad a=c \, , \quad \,
 J_1= J_2=0
\label{c4}
\ee
Series C) contains the Yang--Mills multiplet for $a=0$, $b=1$ and the
K--K tower of short multiplets for $b=p>1$ and $a=c=0$.}
\end{itemize}

\noindent The discussion of the properties of the G--analytic
superfield $W^{12}$ above applies to any of its powers
\begin{equation}\label{e73}
 (W^{12})^p = C[0,p,0]
\end{equation}
(or to $\mbox{Tr}(W^{12})^p$ in the non-Abelian case). The
notation indicates the  Dynkin $SU(4)$ labels $[0,p,0]$ of the
first component of the superfield.

These superfields satisfy the set of G--analyticity conditions
\begin{equation}\label{e6'}
  (\Doa,\ \Dta,\  \bDha,\ \bDfa) C[0,p,0]=0\;,
\end{equation}
and are therefore short multiplets (maximal spin 2= $(1,1)$). As before,
the harmonic conditions
\begin{equation}\label{e61'}
D_J^{\, I} C[0,p,0]=0\;, \qquad I,J=1,2,3,4,\ \ I<J
 \end{equation}
ensure irreducibility under $SU(4)$. Indeed, consider the leading
component
\begin{eqnarray}
[(W^{12})^p]|_0   &=&\phi^{(i_1\ldots i_p)(j_1\ldots
j_p)}u^1_{i_1}\ldots u^1_{i_p} u^2_{j_1}\ldots u^2_{j_p} \quad
\stackrel{D_2^{\,1}}{\rightarrow} \nonumber\\ &&\phi^{(i_1\ldots
i_pj_1)(j_2\ldots j_p)} u^1_{i_1}\ldots u^1_{i_{p}}u^1_{j_1}
u^2_{j_2}\ldots u^2_{j_{p}} = 0\;.  \label{e72b}
\end{eqnarray}
This condition eliminates the symmetrization between the indices
of the first and second set and thus renders the field
irreducible, belonging to the $SU(4)$ UIR $[0,p,0]$. The
alternative proof of irreducibility makes use of the positive root
or the Casimir argument above.
\\
The series above corresponds to K--K towers of IIB supergravity 
on $AdS_5\times S_5$ \cite{krvn} and it was obtained using the oscillator
method by Gunaydin and Marcus \cite{gm}. Its relation with analytic
superfields with harmonic variables of $\o{SU(4)}{SU(2)\times SU(2)\times
U(1)}$ \cite{HW} was discussed in \cite{AF}. 

Another way of obtaining new short representations is to multiply
two $W$'s with different G--analyticities, e.g.
\begin{equation}\label{e74}
[W^{12}(\theta_3,\theta_4,\bar\theta^1, \bar\theta^2)]^{p+q} 
[W^{13}(\theta_2,\theta_4,\bar\theta^1,\bar\theta^3)]^q = C[q,p,q] 
\end{equation}
(we postpone the discussion of the role of the traces to the next
section, and from now on we omit the traces). The lowest component
of the corresponding irreducible superfield belongs to the UIR
$[q,p,q]$  with $p+2q$ indices 1, $p+q$ indices 2 and $q$ indices
3 in the corresponding rows of the YT. (Note that interchanging
$W^{12}$ and $W^{13}$ would give an equivalent series). It is
clear that such superfields satisfy only a subset of the
G--analyticity conditions above:
\begin{equation}\label{e75}
  D^1_\alpha C[q,p,q] = \bar
D^{\dot\alpha}_4 C[q,p,q]  = 0
\end{equation}
and thus depend on 12 out of the total of 16 $\theta$'s. As a
consequence, the value of the spin in their expansions cannot
exceed $3=({3\over 2},{3\over 2})$. Next, we have to impose the
set of H--analyticity constraints
\begin{equation}\label{e78}
D^{\,I}_J C[q,p,q] = 0 \;, \qquad I,J=1,2,3,4,\ \ I<J
\end{equation}
which are clearly compatible with the G--analyticity of
$C[q,p,q]$. Note that they coincide with those for the preceding
series (\ref{e61'}) which is needed for consistency if we set
$q=0$. As before, the effect of these conditions is to single out
the $SU(4)$ UIR $[q,p,q]$. Indeed, the leading component
\begin{equation}\label{e79}
  \phi^{(i_1\ldots i_{p+2q})(j_1\ldots j_{p+q})(k_1\ldots
k_q)}u^1_{i_1}\ldots u^1_{i_{p+2q}} u^2_{j_1}\ldots u^2_{j_{p+q}}
u^3_{k_1}\ldots u^3_{k_q}
\end{equation}
becomes irreducible after imposing the constraints involving
$D_2^{\, 1}\;,\ D_3^{\, 1}\;,\ D_3^{\, 2}$ (they remove all
possible symmetrizations among the different sets of indices).
The above series corresponds to the shortening C) in \eq{c4}.

The third possibility corresponds to the shortening B) in \eq{b4}.
It involves all the three G--analytic $W$'s:
\begin{equation}\label{e80}
 C[q+2n,p,q]= [(W^{12})^{p+q+n}(W^{13})^{q+n}(W^{23})^n]~.
\end{equation}
This time there is only one G--analyticity condition left,
\begin{equation}\label{e81}
  \bar{D}_{4\dot\alpha} C[q+2n,p,q]=0~.
\end{equation}
Consequently, the superfield depends on 14 $\theta$'s (8
left-handed and 6 right-handed) and the spins in its expansion can
go up to $7/2=(2,{3\over 2})$. The H--analyticity constraints are
\begin{equation}\label{e83}
D^{\,I}_J C[q+2n,p,q] = 0 \;, \qquad I,J=1,2,3,4,\ \ I<J\;.
\end{equation}
Once again, they look the same as those for the $C[0,p,0]$ series
(\ref{e61'}) and for the $C[q,p,q]$ series (\ref{e78}) (needed for
consistency). The irrep corresponding to the leading component now
is $[q+2n,p,q]$.

Concluding this subsection we note that there exist other
G--analytic subspaces involving 10 out of the 16 $\theta$'s, for
example, $\theta_{1,2}\;, \bar\theta^{1,2,3}$. However,
superfields living in such subspaces cannot be obtained by
multiplying $W$'s.

\subsubsection{Extra shortening of N=4 superfields}

As in the cases $N=2,3$ before, the $N=4$ representations can in
some case be semishort. The simplest example is the superfield
$W^{12}$ itself, which is ultrashort.
Due to the G-- and H--analyticity constraints
(\ref{e6}), (\ref{e61}) it describes the on-shell $N=4$ ultrashort
SYM multiplet containing six scalars $\phi^{ij}=-\phi^{ji}$, four
spinors $\lambda^{\alpha i}$ (and their conjugates
$\bar\lambda^{\dot\alpha}_i$) and the field strength
$F^+_{\alpha\beta},\ F^-_{\dot\alpha\dot\beta}\;$. Thus, its
$\theta$ expansion is effectively shorter than that of a generic
G--analytic superfield of the type (\ref{e8}).

\begin{eqnarray}
&&W^{12} = \phi^{12} + \theta_3^\alpha \lambda_{4\alpha} -
\theta_4^\alpha \lambda_{3\alpha}+ \bar\theta^{1\da}
\bar\lambda^2_\da - \bar\theta^{2\da} \bar\lambda^1_\da+
\theta_3^\alpha\theta_4^\beta F^+_{\alpha\beta} +
\bar\theta^{1\da} \bar\theta^{2\db} F^-_{\da\db} +\nonumber\\ &&+i
\theta_3^\alpha\btob\padb\phi^{23} +i
\theta_4^\alpha\btob\padb\phi^{24} - i
\theta_3^\alpha\bttb\padb\phi^{13} -i
\theta_4^\alpha\bttb\padb\phi^{14} \nonumber\\ &&+
i\bar\theta^{1\db}\theta^{(\alpha}_3\theta^{\beta)}_4\padb\lambda_{1\beta}
+i\bar\theta^{2\db}\theta^{(\alpha}_3\theta^{\beta)}_4\padb\lambda_{2\beta}
+i\theta^\alpha_3\bar\theta^{1(\da}\bar\theta^{2\db)}\padb\bar\lambda^3_\db
\nn\\&&
+i\theta^\alpha_4\bar\theta^{1(\da}\bar\theta^{2\db)}\padb\bar\lambda^4_\db
 +\theta^{(\alpha}_3\theta^{\beta)}_4\bar\theta^{1(\da}\bar\theta^{2\db)}
\pada\partial_{\beta\db} \phi^{34}~. \label{e66}
\end{eqnarray}
To see this take, for instance, the component at
$\theta_3\bar\theta^1$. It can be defined using the spinor
derivatives of $W^{12}$ at $\theta=0$, $A^{23}_\adb = D^3_\alpha
\bar D_{1\db} W^{12}\vert_0$. Now, $W^{12}$ is subject to the
H--analyticity conditions (\ref{e61}), in particular, $D_3^{\,
1}W^{12}=0$. Applying this to the component and using the
G--analyticity condition $D^1_\alpha W^{12}=0$, we find
\begin{equation}\label{e65}
 D_3^{\, 1}A^{23}_\adb =
  D^1_\alpha\bar D_{1\db} W^{12}\vert_0 =i\padb
 W^{12}\vert_0 = i\partial_\adb \phi^{12}  \;.
\end{equation}
The resulting harmonic equation has the obvious solution
$A^{23}_\adb = -i\partial_\adb \phi^{23} $.

Inspecting the superfield expansion (\ref{e66}) one sees that each
$\theta$ (or $\bar\theta$) appears only linearly. This means that
the superfield $W^{12}$ satisfies Grassmann linearity conditions
of the type, e.g.,
\begin{equation}\label{e67}
  (D^3)^2 W^{12}=0\;.
\end{equation}
Once again, this constraint can be easily derived by using the
basic G-- and H--analyticity properties of  $W^{12}$. Indeed,
denote $A^{1233} = (D^3)^2 W^{12}$ and hit it with the harmonic
derivatives $ D_3^{\, 1}\;,\  D_3^{\, 2}$:
\begin{equation}\label{e68}
  D_3^{\, 1}A^{1233} = D_3^{\, 2}A^{1233}=0\;.
\end{equation}
These two constraints ensure the $SU(4)$ reducibility of  (the
leading component of) $A^{1233}$ by eliminating all the
symmetrizations between indices projected with different
harmonics. Then we can rewrite it as $A^{1233}= A^3_4$ and by
hitting it with $D_3^{\, 1}$ we find
\begin{equation}\label{e69}
  D_3^{\, 1}A^3_4 = A^1_4 = 0 \ \Rightarrow \ A^{1233} = 0\;.
\end{equation}

In the same way one can readily prove the following relations
\be
(D^3D^4)W^{12}=(\bar{D}_1\bar{D}_2)W^{12}=0\;, \ee
\be
(D^3)^2 W^{12}= (D^4)^2 W^{12}=(\bar{D}_1)^2 W^{12}=(\bar{D}_2)^2
W^{12}= 0~. \ee

Another way to find out that the superfield $W^{12}$ is
ultrashort is to notice that the components
\be
F^+_{\alpha\beta}=D^3_{(\alpha}D^4_{\beta)} W^{12}~,\qq
F^-_{\dot\alpha\dot\beta}=\bar{D}_{1(\da} \bar{D}_{2\db)} W^{12}
\ee
are the highest (in this case the only) chargeless ones in the
expansion. Adding one more spinor derivative ({\ie}, moving a step
up in the expansion) produces a charged component which is
eliminated by the harmonic conditions. For example, take $\psi^3=
D^3\bar D_1\bar D_2 W^{12}\vert_0$ and hit it with  $D_3^{\, 1}$:
$$
  D_3^{\, 1}\psi^3_{\alpha\dot\alpha\dot\beta} =
D_3^{\, 1}\Dha\bDoa\bDtb W^{12}\vert_0 = -i\pada \bDtb
W^{12}\vert_0 = -i\pada \bar\chi^1_{\dot\beta} $$
\begin{equation}\label{e70}
  \Rightarrow\ \ \psi^i_{\alpha\dot\alpha\dot\beta}
 =-i\pada \bar\chi^i_{\dot\beta} \;.
\end{equation}
Thus, we can say that the expansion of the superfield $W^{12}$
ends at the level of 2 $\theta$'s (in the sense that all the
higher components are expressed in terms of derivatives of the
lower ones). We call such superfields ultrashort.

The linearity property of $W^{12}$ is of course lost when we start
multiplying them. Nevertheless, $(W^{12})^2$ and  $(W^{12})^3$ are
still shorter than a generic superfield of the same G--analyticity
type. According to the general discussion in section 2, $p=2$
gives a ``current'' superfield while $p=3$ gives a semishort
multiplet.
 Indeed, let us examine the top component of
$C[0,p,0]\sim (W^{12})^p$:
\begin{equation}\label{e90}
 \Phi^{(0,p-4,0)}~ \sim~ (D^3D^4\bar D_1\bar D_2)^2
(W^{12})^p\vert_{\theta=0} \;.
\end{equation}
For $p\geq 4$ this is a field containing an $SU(4)$ irrep which
survives all the harmonic conditions (for $p=4$ it becomes a
singlet), so the superfield is not ultrashort. For $p=3$ we find
singlets at the level of 6 $\theta$'s:
\begin{equation}\label{??}
 (D^3)^2(D^4)^2 \bar{D}_{1(\da} \bar{D}_{2\db)}C[0,3,0]~, \qquad
D^3_{(\alpha}D^4_{\beta)}(\bar{D}_1)^2(\bar{D}_3)^2 C[0,3,0]~.
\end{equation}
They are not affected by the harmonic conditions and indeed, by
taking the expansion (\ref{e66}) of a single $W^{12}$ to the third
power we do find such components. Any higher component will carry
a charge and will be killed by the harmonic conditions (for
instance, $\psi^4 =  D^3 (D^4\bar D_1\bar D_2)^2
(W^{12})^3\vert_{\theta=0} $ is annihilated by $D_4^{\,1}$). Thus,
the expansion of $C[0,3,0]$ ends at 6 $\theta$'s.

Similarly, for $C[0,2,0]=(W^{12})^2$  we find the singlets \bea
&&(D^3)^2(D^4)^2   C[0,2,0]
 ~,\nn\\ &&(\bar{D}_1)^2(\bar{D}_3)^2 C[0,2,0]      ~,\nn\\
&&D^3_{(\alpha}D^4_{\beta)} \bar{D}_{1(\da} \bar{D}_{2\db)}
C[0,2,0]      ~,\nn \eea
which are indeed present in the square of the expansion
(\ref{e66}). Thus, $C[0,2,0]$ is another extrashort superfield
ending at 4 $\theta$'s, it is in fact a ``current superfield''.

In the cases of the series $C[q,p,q]$ with $q>1$ and $C[q+2n,p,q]$
with $n>1$ the same analysis shows that the top component is
always present.
In the case $(W^{12})^{p+1}W^{13}$ the superfield is linear in
$\theta_2$ because this is the property of $W^{13}$ and $W^{12}$
does not depend on $\theta_2$. Similarly, the product
$(W^{12})^{p+q+1}(W^{13})^{q+1}W^{23}$ is linear in $\theta_1$.
The above cases correspond to semishortening.
These superfields are even shorter for certain values of $p$ and
$q$, but this requires an additional analysis.


\section{Multitrace operators and multiparticle states}

The analysis of different classes of $N=4$ conformal supermultiplets
obeying different types of shortening conditions has an interesting
application to some ``states'' which are not K--K states but have
rather the interpretation of ``multiparticle states'' \cite{agmo,cs}
in the AdS/CFT correspondence.

In $N=4$ Yang--Mills theory these states correspond to the decomposition
of the product of some ``short'' (single trace)
 K--K multiplets into irreducible
superconformal blocks.
Such blocks necessarily contain multitrace (rather than single trace)
\cite{hfmmr,hmmr2,cs}
Yang--Mills gauge-invariant operators which in general are not in the
same representations of the ``short'' K--K multiplets.

In this section we will make the analysis for the most general
double-trace and triple-trace gauge invariant operators of $N=4$
Yang--Mills. The extension to higher multitraces is in principle
straightforward.

Let us denote by $\phi^{12},\phi^{13},\phi^{23}$ the lowest
$\theta$ components of the three superfields
$W^{12},W^{13},W^{23}$. Consider now the gauge invariant
operators:
\begin{eqnarray}
s) && \trace [(\phi^{12})^p] \nonumber \\
d) &&\trace [(\phi^{12})^{p+q}]\trace [(\phi^{13})^q] \nonumber \\
t) && \trace [(\phi^{12})^{p+q+n}]\trace [(\phi^{13})^{q+n}]
\trace [(\phi^{23})^n]
\end{eqnarray}
Sequence $s)$ is the usual K--K tower of IIB supergravity on
$AdS_5 \times S_5$. It gives all multiplets with $J_{max}=2$
(more precisely a $(1,1)$ tensor in the $[0,p-2,0]$ UIR of $SU(4)$)
and the first component is
\be
\trace (\phi^{\ell_1} \cdots \phi^{\ell_p}) \quad \mbox{symmetric traceless}
\label{single}
\ee
where $\phi^\ell$ is a scalar in the $[0,1,0]$ of $SU(4)$ defined as:
\be
\phi^\ell = W^\ell |_{\theta =0} \equiv (\gamma^\ell )_{ij} W^{ij}.
\ee
Sequences $d)$ are the double-trace operators.
The $\theta =0$ term is contained in the product
\be
\trace (\phi^{\ell_1} \cdots \phi^{\ell_{p+q}})
 \trace (\phi^{m_1} \cdots \phi^{m_q})
\label{double}
\ee
where each single trace is symmetric traceless.

As an illustrative example consider for instance the product of two
lowest components of the ``current'' multiplet:
\be
\trace \left[\phi^{(\ell_1}\phi^{\ell_2)}-\o 16 \epsilon^{\ell_1\ell_2}
\phi^{(m}\phi_{m)}\right] 
\trace \left[\phi^{(\ell_3}\phi^{\ell_4)}-\o 16 \epsilon^{\ell_3\ell_4}
\phi^{(n}\phi_{n)}\right] 
\ee
It contains the irreducible $SU(4)$ components:
\be
\begin{array}{ccccccccc}
20 \times  20 &=& 105 & +& 84 &+& 20 &+& 1 \cr
&& [0,4,0] && [2,0,2] && [0,2,0] && [0,0,0]
\end{array}
\ee
The $(105)$ and $(84)$ correspond to two short multiplets
with top spin $(1,1)$ and $(\o 32 , \o 32)$ respectively, while the last
two are long multiplets with top spin $(2,2)$.
The latter acquire anomalous dimensions in perturbation theory as shown
in refs. \cite{cs}-\cite{bkrs}.
\\
The first two UIRs are contained in 
$
\trace \left(\phi^{12}\right)^2 \trace \left(\phi^{12}\right)^2 
$ and 
$
\trace \left(\phi^{12}\right)^2 \trace \left(\phi^{13}\right)^2 
$ respectively, while the last two correspond to UIRs superfields with
$8\theta$, $8\bar\theta$.

The multiplets in \eq{double}decompose in long ($J_{max} =4$) and
short multiplets ($J_{max}=2,3$). The virtue of the multiplication
in $d)$ is that precisely it singles out all shortening occurring
in \eq{double}.

Note that the operator
\be
\trace (\phi^{12})^\ell \trace (\phi^{12})^m \ee gives the same
UIR as $\trace (\phi^{12})^{\ell +m}$ {\ie}  ${\mathcal D}(\ell
+m,0,0 ; 0,\ell +m,0)$. This means that the $[0,p,0]$ UIR obtained
in any multitrace operator is a short multiplet. The same type of
argument will apply to the other shortenings.

Analogously, an operator of the type
\be
\trace \left([\phi^{12}]^\ell [\phi^{13}]^m\right)
\ee
would correspond to a single trace first component ($\theta =0$) operator:
\be
\trace \phi^{a_1}\cdots \phi^{a_\ell}\phi^{b_1}\cdots \phi^{b_m}
\ee where some antisymmetrization of two $a,b$ indices occurs so
it would not be a superconformal primary operator in super
Yang--Mills theory \cite{wit} (although it would be conformal primary in an
Abelian theory where traces are removed, since in that case it
would coincide with
 $(\phi^{12})^\ell (\phi^{13})^m$).
Also note that in a rank 1 Abelian theory only s) would survive
because in that case all antisymmetrizations of $\phi^{\ell}$
would vanish. These considerations also imply that no single power
of any $\phi^{12}$ (or $\phi^{13}$, $\phi^{23}$) should occur in
the product (since $\trace \phi =0$) then implying that linear semishort
operators ({\ie}  operators satisfying a $D^2 =0$ constraint) do
not occur in $SU({\mathcal N})$ Yang--Mills theory.

{}From the general class of shortening we see that $d)$ contains
the irreducible pieces which correspond to the shortening of:
\be
\trace [(\phi^{12})^{p+q+k}]\times \trace [(\phi^{13})^{q-k}] \, ,
\quad 0\leq k \leq q \ee {\ie} to highest weight states of the
type  ${\mathcal D}(2q+p,0,0;q-k,p+2k,q-k)$, the new one being the $k=0$ one,
with Dynkin label $[q,p,q]$ ($k=q-1$ is missing because $ \trace
\phi^{13}=0$).
This is precisely the UIR singled out by the H-analyticity constraints
\eq{e78}

All these states have quantized dimensions and lie in multiplets with
$J_{max}=3$, unless $k=q$, for which $J_{max}=2$.


\begin{figure}[h]
\setlength{\unitlength}{2mm}
 \begin{center}
\label{doublerep}
 \caption{$[q,p,q]$ representation:}
\begin{picture}(8,8)
\put(0,0){\framebox (3,3){\tiny $1_1$}}
\put(3,0){\framebox (12,3){\tiny $\cdots$}}
\put(15,0){\framebox (3,3){{\tiny $1_{p+2q}$}}}
\put(0,-3){\framebox (3,3){\tiny $2_1$}}
\put(3,-3){\framebox (8,3){\tiny $\cdots$}}
\put(11,-3){\framebox (3,3){\tiny $2_{p+q}$}}
\put(0,-6){\framebox (3,3){\tiny $3_{1}$}}
\put(3,-6){\framebox (4,3){\tiny $\cdots$}}
\put(7,-6){\framebox (3,3){\tiny $3_{q}$}}
\end{picture}
  \end{center}
\end{figure}
\vspace{0.5cm}

\noindent
Let us apply this to the cases of double-traces with $d\leq 6$.
The rational to restrict to $s)$, $d)$, $t)$ families is the following:
\begin{itemize}
\item{
 $\trace [(\phi^{12})^2]\times \trace [(\phi^{13})^2]$.

In this case $p=0$, $q=2$, so the two short reps. are the $[0,4,0]$
($J_{max}=2$) and $[2,0,2]$ ($J_{max}=3$).
It has been confirmed by direct calculation that indeed two such objects
are not renormalized in perturbative theory (at one loop).
}
\item{
$\trace [(\phi^{12})^3]\times \trace [(\phi^{13})^2]$.

In this case $p=1$, $q=2$. The short multiplets are in the $[0,5,0]$ and
$[2,1,2]$.
}
\item{
$\trace [(\phi^{12})^3]\times \trace [(\phi^{13})^3]$.

In this case $p=0$, $q=3$. The short multiplets are in the $[0,6,0]$,
$[3,0,3]$ and $[2,2,2]$.
}
\item{
$\trace [(\phi^{12})^4]\times \trace [(\phi^{13})^2]$.

In this case $p=2$, $q=2$. The short multiplets are in the $[0,6,0]$ and
$[2,2,2]$.
}
\end{itemize}

It is thus obvious that the number of short multiplets is precisely
$q$ of which $q-1$ have $J_{max}=3$ and one has $J_{max}=2$.

We now consider triple-trace operators where a new type of shortening
($J_{max}=\frac{7}{2}$) occurs.

The generic triple-trace operator is:
\be
\trace [(\phi^{12})^{p+q+n}]\trace [(\phi^{13})^{q+n}]\trace [(\phi^{23})^n]
\ee
with dimension $d=p+2q+3n$.
\\
The above expressions single out the short multiplets contained in the following
triple-trace operator composites:
\be
\trace (\phi^{\ell_1} \cdots \phi^{\ell_{p+q+n}}) \trace
(\phi^{m_1} \cdots \phi^{m_{q+n}}) \trace (\phi^{s_1} \cdots
\phi^{s_n}) \label{triple} \ee The new phenomenon here is that
three types of short multiplets with $J_{max} = 2,3$ and
$\frac{7}{2}$ occur. The new short multiplet is a ``chiral''
superfield whose first component is in the $[q+2n,p,q]$ of $SU(4)$
and with a $J_{max}=\frac{7}{2}$ state in the $(2,\frac{3}{2})$
rep. of $SL(2,\IC)$.
This is the UIR singled out by the constraints \eq{e83}
{}From the analysis of
the product of:
\be
\trace [(W^{12})^{p+q+n}]\trace [(W^{13})^{q+n}]Tr[(W^{23})^n]
\ee
it follows that the above triple-trace operators contain all the shortenings
which occur in:
\be
\trace [(\phi^{12})^{p+q+n+x}]
\trace [(\phi^{13})^{q+n+k-x}]
\trace [(\phi^{23})^{n-k}] \, ,
\quad  0\leq k \leq n , \, 0 \leq x \leq q+2k
\ee
 if $k \leq p$, or
\be
\trace [(\phi^{12})^{q+k+n+x}]
\trace [(\phi^{13})^{p+q+n-x}]\trace [(\phi^{23})^{n-k}] \, ,
\quad  0\leq k \leq n , \, 0 \leq x \leq p+q+k
\ee
for $k \geq p$.\\

\begin{figure}[h]
\setlength{\unitlength}{2mm}
 \begin{center}
\label{triplerep1}
 \begin{picture}(6,6)
\put(0,0){\framebox (30,3){$p+2q+2n+k$}}
\put(0,-3){\framebox (25,3){$p+q+2n-k+x$}}
\put(0,-6){\framebox (20,3){$q+2n-x$}}
\end{picture}
\vskip 1 cm
\caption{$[q+2n-x,p-k+2x,q+2k-x]$ representation,
 for $k\leq p$; $0\leq k \leq n$,
$0\leq x \leq q+2k$:}
 \end{center}
\end{figure}

\begin{figure}[h]
\setlength{\unitlength}{2mm}
 \begin{center}
\label{triplerep2}
 \begin{picture}(6,6)
\put(0,0){\framebox (30,3){ $p+2q+2n+k$}}
\put(0,-3){\framebox (25,3){ $q+2n+x$}}
\put(0,-6){\framebox (20,3){ $p+q+2n-k-x$}}
\end{picture}
\vskip 1cm
\caption{$[p+q+2n-k-x,-p+k+2x,p+q+k-x]$ representation,
 for $k\geq p$; $0\leq k \leq n$,
$0\leq x \leq p+q+k$:}
 \end{center}
\end{figure}

The first triple trace operator is ($d=6$):
\be
\trace [(\phi^{12})^2]\trace [(\phi^{13})^2]\trace [(\phi^{23})^2] \, ,
\quad (p=q=0 ; n=2 )
\ee
For $k=0$, $x=0$ it gives the $[0,0,4]+[4,0,0]$ ($J_{max}=\frac{7}{2}$).
For $k=2$, it gives all the shortenings already occurred in
$\trace [(\phi^{12})^{4+x}]\trace [(\phi^{13})^{2-x}]$.
These are the $[0,6,0]$ and $[2,2,2]$ ($J_{max}=2$ and $3$ respectively).

This completes the analysis of the shortenings of double and triple
trace operators.
Of course, due to their ring structure, higher multiple trace operators 
can be obtained by further multiplying structures as in s) d) t) then 
obtaining the same type of shortening as in the previous composite
operators.

The bulk interpretation of these composite operators is that there are some
multiparticle BPS states in the supergravity side.

The ``non-renormalization'' of the $[0,4,0]$ and $[2,0,2]$ short multiplets
contained in the two graviton-multiplets particle state was shown 
in $N=4$ Yang--Mills perturbation theory in ref.
\cite{bkrs,sk}. The latter reference extended the analysis for the
$(0,p,0)$ block to all multitrace components.
Its relation with shortening was established in ref. \cite{AF} and \cite{FZ}.

{}From the shortening conditions we see that while the usual K--K
states are $\o 12$ BPS (since the superfield does not depend on
 4 $\theta_L$, 4 $\bar\theta_R$) the new
short classes correspond to $\o 14$ BPS (2 $\theta_L$, 2 $\bar\theta_R$)
and $\o 18$ BPS (2 $\bar\theta_R$).
The lowest dimensional $\o 14$ BPS operators occur for highest weight
${\mathcal D} (4,0,0;2,0,2)$ (double-trace) while the lowest dimensional
$\o 18$ BPS states occur for highest weights
 ${\mathcal D} (6,0,0;4,0,0)$ + c.c..

\setcounter{equation}0
\section{Conclusion}
In the present paper we have analyzed all possible shortenings which
are obtained by composite operators made out of the field strength gauge
multiplets and hypermultiplets (for $N=2$).

These shortenings are characterized by subspaces of HSS's which do
not depend on a certain number of (fermionic) $\theta$ variables.

If a certain subspace (of the full superspace with $4N$ $\theta$ variables)
does not depend on $n$ fermionic coordinates, then a superfield on such a
space is generally called $\o{n}{4N}$ BPS in analogy with a particle state
interpretation.

Moreover, if $n=n_L+n_R$ then the highest spin of such a superfield is
\be
(J_1,J_2)=\left(\o{2N-n_L}{4},\frac{2N-n_R}{4}\right). \ee We can
summarize the set of subspaces, for each $N$, by the pair
$(n_L,n_R)$.

All possible shortenings found from the analysis of section 3 are
summarized in the following table.

\begin{table}[h]
  \begin{center}
    \leavevmode
\label{bps}
    \begin{tabular}{cccc}
$N$ & $(n_L,n_R)$ & $(J_1,J_2)$ & BPS \\
\hline
      & $(2,2)$ & $(\o 12,\o 12)$ & $\o 12$ \\
$N=2$ & $(0,4)$ & $(1,0)$ & $\o 12$  \\
      & $(0,2)$ & $(1,\o 12)$ & $\o 14$  \\
\hline
      & $(2,4)$ & $(1,\o 12)$ & $\o 12$ \\
$N=3$ & $(2,2)$ & $(1,1)$ & $\o 13$  \\
      & $(0,2)$ & $(\o 32,1)$ & $\o 16$  \\
\hline
      & $(4,4)$ & $(1,1)$ & $\o 12$ \\
$N=4$ & $(2,2)$ & $(\o 32 ,\o 32)$ & $\o 14$  \\
      & $(0,2)$ & $(2,\o 32)$ & $\o 18$  \\
    \end{tabular}
  \end{center}
\end{table}

All these representations refer to UIR's with highest weight
states with $J_1=J_2=0$. In this case the generic ``long
multiplets'' (massive non-BPS states) have $J_1=J_2=1,\o 32 ,2 $
for $N=2,3,4$ respectively.

Short multiplets have ``protected dimensions'' in conformal field theories.
This is not the case for long multiplets whose dimension is then renormalized.

We found, as an application of these results, that some multiparticle
state channels, occurring in $AdS_5\times S_5$ compactifications of IIB
string theory, correspond to such short representations.

The AdS/CFT correspondence of supergravity with large ${\mathcal N}$ gauge
theories then predicts that supergravity correlators in these channels
would exhibit ``canonical dimensions'', then implying a new kind of
``non-renormalization theorems'' for $N=4$ super Yang--Mills theory.

\section*{Acknowledgements}

E.S. and B.Z. are grateful to E.A. Ivanov for helpful discussions.
B. Zupnik thanks P. Sorba for the kind hospitality extended to him
during his stay at LAPTH.
S.F. would like to thank M. Bianchi, M. Gunaydin, R. Stora
and A. Van Proeyen for
enligthening conversations.
 The work of B.Z. is partially supported
by the grants  RFBR-99-02-18417, RFBR-DFG-99-02-04022,
INTAS-96-0308 and NATO-PST.CLG-974874. The work of L.A. and S.F.
has been supported by the European Commission TMR programme
ERBFMRX-CT96-0045.
S.F. is also supported in part by DOE grant DE-FG03-91ER40662, Task C.

\appendix
\setcounter{equation}0

\section{\lb{E} Appendix }
\subsection{Harmonic variables}

We introduce harmonic variables on the coset $SU(N)/U(1)^{N-1}$ in
the form of $SU(N)$ matrices $u_i^I$ or their complex conjugates
$u^i_I$. Here $i$ is an index of the fundamental representation of
$SU(N)$ whereas $I=1,2,\ldots,N$ is an index denoting the set of
$N-1$ $U(1)$ charges. The choice of the charges for $N=2,3,4$ is
as follows:
\begin{eqnarray}
N=2:\  && u^1_i = u^{(1)}_i\;, \ \ u^2_i = u^{(-1)}_i\;;
\nonumber\\ N=3: \ && u^1_i=u^{(1,0)}_i~,\ \ u^2_i=u^{(-1,1)}_i~,
\ \  u^3_i=u^{(0,-1)}_i~;\label{a1} \\  N=4: \  &&
u^1_i=u^{(1,0,1)}_i~, \ \   u^2_i=u^{(-1,0,1)}_i~, \ \
u^3_i=u^{(0,1,-1)}_i~, \ \  u^4_i=u^{(0,-1,-1)}_i~; \nonumber
\end{eqnarray}
the conjugates have the opposite charges, e.g., in $N=2\ $ $u^i_1
= u^{i(-1)},\ u^i_2 = u^{i(1)}$. The fact that the $u$'s form an
$SU(N)$ matrix implies the following constraints:
\begin{eqnarray}
 &&u^I_i u^i_J=\delta^I_J~,\lb{a2} \\u\in
SU(N):\quad &&u^I_i u^k_I =\delta^k_i~,\lb{a2'}\\ &&
\varepsilon^{i_1\ldots i_N}u^1_{i_1}\ldots u^N_{i_N}=1~.\lb{a2''}
\end{eqnarray}
Exceptionally, in the case $N=2$ one can raise and lower the
indices of the harmonics with the help of the Levi-Chivita  symbol
$$
\varepsilon_{ik}\varepsilon^{kl}=\delta^l_i~,\quad
\varepsilon^{12}=-\varepsilon_{12}=1~.
$$
This property allows us to identify the two sets of harmonics
$u_i^I$ and $u^i_I$:
$$
u^1_i\equiv \varepsilon_{ij} u^j_2 \equiv u_{2i}\equiv u^+_i~,\qq
u^2_i\equiv -\varepsilon_{ij} u^j_1 \equiv-u_{1i}\equiv u^-_i~.
$$
Note that in ref. \cite{GIK1}, where the $N=2$ harmonic variables
have been introduced for the first time, the notation $u^\pm_i$
was used. Here we prefer to have an uniform notation valid for any
$N$.

The harmonic functions are supposed to transform homogeneously
under $U(1)^{N-1}$, {\ie} they carry definite $U(1)$ charges. This
means that the dependence on the matrix variables $u$ is
considered modulo  $U(1)^{N-1}$ transformations, which provides
an $SU(N)$ covariant way to parametrize the coset
$SU(N)/U(1)^{N-1}$. These functions are given by their infinite
harmonic expansions on the coset. For instance, the function
$f^1(u)$ will have the following expansion in $N=2$:
\begin{equation}\label{a3}
  f^1(u) = f^iu^1_i + f^{(ijk)}u^1_iu^1_ju^2_k + \ldots
\end{equation}
going over the totally symmetrized multispinors (irreps) of
$SU(2)$. In $N=3$ this expansion is considerably richer,
\begin{equation}\label{a4}
  f^1(u) = f^iu^1_i + f^{(ij)}_k u^1_iu^2_ju^k_2 + g^{(ij)}_k u^1_iu^3_ju^k_3
+ h_{(ij)}u^i_2 u^j_3 +\ldots
\end{equation}
and goes over all possible irreps of $SU(3)$ such that after
projection with $u$'s the total charge will be that of $f^1$.

The harmonic coset  $SU(N)/U(1)^{N-1}$ has $N(N-1)/2$ complex
dimensions. Correspondingly, there are as many covariant
derivatives on it. In our $SU(N)$ covariant description of the
coset these derivatives are made out of the operators
\begin{equation}\label{a5}
  \partial^{\,I}_J = u^I_i{\partial\over\partial u^J_i} -
u^i_J{\partial\over\partial u^i_I}
\end{equation}
which respect the defining constraints (\ref{a2}), (\ref{a2'}).
The third constraint (\ref{a2''}) implies that the charge-like
operators $\partial^{\,I}_I$ are not independent,
\begin{equation}\label{a6}
  \sum_{I=1}^N \partial^{\,I}_I = 0\;.
\end{equation}
These derivatives act on the harmonics as follows:
\be
\partial^I_J u^K_i=\delta^K_J u^I_i~,\qquad
\partial^I_J u^i_K=-\delta^I_K u^i_J~.\lb{F9}
\ee So, the $N-1$ $U(1)$ charges are counted  by the  derivatives
\begin{eqnarray}
N=2:\  && H=\poo -\ptt \;; \nonumber\\ N=3: \ && H =\poo
-\ptt~,\quad H^\prime =\ptt-\phh~;\label{a7} \\  N=4: \  && H
=\poo -\ptt~,\quad H^\prime =\phh-\pff~,\q H^{\prime\prime} =\poo
+\ptt-\phh -\pff~. \nonumber
\end{eqnarray}
The remaining $N(N-1)/2$ derivatives $\partial^{\,I}_J $, $I<J$
(or their conjugates $\partial^{\,I}_J $, $I>J$) are the true
harmonic derivatives on the  coset  $SU(N)/U(1)^{N-1}$.

It is important to realize that the set of $N^2-1$ derivatives
$\partial^{\,I}_J $ (taking into account the linear dependence
(\ref{a6})) form the algebra of $SU(N)$:
\begin{equation}\label{a8}
 [\partial^{\,I}_J,\partial^{\,K}_L ]=\delta^K_J \partial^{\,I}_L -\delta^I_L
\partial^{\,K}_J~.
\end{equation}
The Cartan decomposition of this algebra $L^+ + L^0 + L^-$ is
given by the sets
\begin{equation}\label{a9}
  L^+ = \{\partial^{\,I}_J \;, \ I<J\}\;, \quad
L^0 = \{\partial^{\,I}_I\;, \ \sum_{I=1}^N \partial^{\,I}_I = 0
\}\;, \quad  L^- = \{\partial^{\,I}_J \;, \ I>J\}\;.
\end{equation}
It becomes clear that imposing the harmonic conditions
\begin{equation}\label{a10}
  \partial^{\,I}_J f^{K_1\ldots K_q}(u) = 0\;, \quad I<J
\end{equation}
on a harmonic function with a given set of charges $K_1\ldots K_q$
defines a highest weight of an $SU(N)$ irrep. In other words, the
harmonic expansion of such a function contains only one irrep
which is determined by the combination of charges $K_1\ldots K_q$.
For instance, in $N=2$ we have (see (\ref{a3}))
\begin{equation}\label{a11}
  \partial^1_2 f^1(u)= 0 \ \Rightarrow \  f^1(u) = f^iu^1_i\;,
\end{equation}
and, similarly, in $N=3$  (see (\ref{a4}))
\begin{equation}\label{a12}
 \partial^1_2 f^1(u)= \partial^1_3 f^1(u)=\partial^2_3 f^1(u)=0 \
\Rightarrow \  f^1(u) = f^iu^1_i\;.
\end{equation}
Note that not all of the conditions (\ref{a12}) are independent
since $\partial^1_3 = [\partial^1_2,\partial^2_3]$. Written down
in a complex parametrization of the coset, conditions (\ref{a10})
take the form of harmonic (H-)analyticity conditions  on the
function $f(u)$. It is important to realize that for certain
combinations of charges the condition (\ref{a10}) may not have a
non-trivial solution. For example, the function $f^2(u)$ cannot be
H-analytic since $\partial^1_2 f^2(u) = f^iu^1_i = 0 \ \Rightarrow
\ f^i=0$.

\subsection{Grassmann analyticity}

The introduction of harmonic coordinates allows one to define
various subspaces of the full $N$-extended superspace involving
only a subset of the Grassmann coordinates without breaking
$SU(N)$. Indeed, we can rewrite the supersymmetry transformations
in terms of the harmonic-projected Grassmann variables as follows:
\begin{eqnarray}
  &&\delta x^\adb =i(\theta^\alpha_I u^I_i\bar{\epsilon}^{\db i} -
 \epsilon^\alpha_i u^i_I\bar{\theta}^{\db I} )~, \nonumber\\
  &&\delta\theta^\alpha_I =\epsilon^\alpha_i u^i_I~,  \label{a13}\\
  &&\delta\bar{\theta}^{\db I} = u^I_i\bar{\epsilon}^{\db i}\nonumber
\end{eqnarray}
where $\theta^\alpha_I = \theta^\alpha_i u^i_I,\ \bar{\theta}^{\db
I} = u^I_i\bar{\theta}^{\db i}$. Now, we can shift $x^\adb$ in a
variety of ways such that the transformation of the new variable
does not involve some of the projections of $\theta$ or
$\bar\theta$. Thus we obtain subspaces of the full superspace
closed under supersymmetry. Such superspaces are called Grassmann
(G-)analytic. Here are some examples:
\begin{eqnarray}
N=2:\ && x_\A^\adb= x^\adb
+i(\theta_2^{\alpha}\bar{\theta}^{2\db}-\theta_1^{\alpha}
\bar{\theta}^{1\db})~, \ \theta_2^{\alpha}~, \
\bar{\theta}^{1\da}~;\nonumber\\ N=3:\ &&x^\adb_\A =x^\adb
+i(\tta\bttb+\tha\bthb- \toa\btob )~,\ \tta~,\ \tha~,\
\bar{\theta}^{1\da}~; \label{a14}\\
 N=4:\   &&x^\adb_\A=x^\adb + i(\tha\btha
+\tfa\btfa - \toa\btob - \tta\bttb)~, \  \theta_3^{\alpha}~, \
\theta_4^{\alpha}~, \ \btoa~, \ \btta~. \nonumber
\end{eqnarray}
In these examples the G-analytic superspace has the minimal odd
dimension possible, {\ie} half of the total number $4N$. In this
sense the G-analytic superspaces are analogs of chiral superspace,
which also involves the left- or right-handed half of the odd
variables. However, an important difference is that in the cases
$N>2$ one can also have G-analytic superspaces with intermediate
odd dimensions, {\ie} 8 and 10 in $N=3$ and 10, 12 and 14 in
$N=4$. The reason is that the harmonics on the coset
$SU(N)/U(1)^{N-1}$  allow one to break the spinor variables up
into  $N$ independent projections, whereas the chiral projection
always picks a spinor in the fundamental representation of
$SU(N)$.

An equivalent definition of G-analyticity is to consider
superfields satisfying constraints involving the spinor
derivatives $D^i_\alpha$ and $\bar D_{i\dot\alpha}$.
These derivatives commute with supersymmetry and satisfy the
following algebra \bea &&\{ D^k_\alpha, D^l_\beta\}=0~,\nonumber\\
&&\{ \bar{D}_{k\da}, \bar{D}_{l\db}\}=0~,\lb{a16}\\ &&\{
D^k_\alpha, \bar{D}_{l\db}\}= i\delta^k_l\padb~.\nonumber \eea
which resembles the supersymmetry algebra. Now, projecting them
with harmonics, we can impose a number of G-analyticity conditions
on the superfields $\Phi(x,\theta,\bar\theta)$. For example, the
conditions corresponding to the subspaces (\ref{a14}) are
\begin{eqnarray}
N=2:\ && D^1_\alpha\Phi = \bar D_{2\dot\alpha}\Phi=0~;\nonumber\\
N=3:\ && D^1_\alpha\Phi = \bar D_{2,3\dot\alpha}\Phi=0~;
\label{a17}\\
 N=4:\   && D^{1,2}_\alpha\Phi = \bar D_{3,4\dot\alpha}\Phi=0~. \nonumber
\end{eqnarray}
It is clear that this can be done with any subset of $D$'s and
$\bar D$'s as long as they anticommute.

The role of the shifts of $x$  in  (\ref{a14}) is to define a
G-analytic  basis in which the derivatives in (\ref{a17}) become
torsion-free, e.g.
$$
D^1_\alpha = \partial^1_\alpha\;, \quad \bar D_{2\dot\alpha} =
\bar\partial_{2\dot\alpha}\;, \quad \mbox{etc.}
$$
Of course, the spinor derivatives which do not belong to the
analytic set still involve space-time derivatives in this basis.
More important, the harmonic derivatives acquire torsion terms in
the G-analytic basis. Thus, in the bases (\ref{a14}) one has
\begin{equation}\label{a18}
 \Doh =\poh -i \tha\btob\padb - \tha\poa +\btoa
\bpha\;, \quad \mbox{etc.}
\end{equation}
This implies that the condition of H-analyticity on harmonic
superfields $\Phi(x_A,\theta,\bar\theta,u)$ involves space-time
derivatives of the components.

\end{document}